# Agora Elevator Bodily Sensation Study

**REPORT**

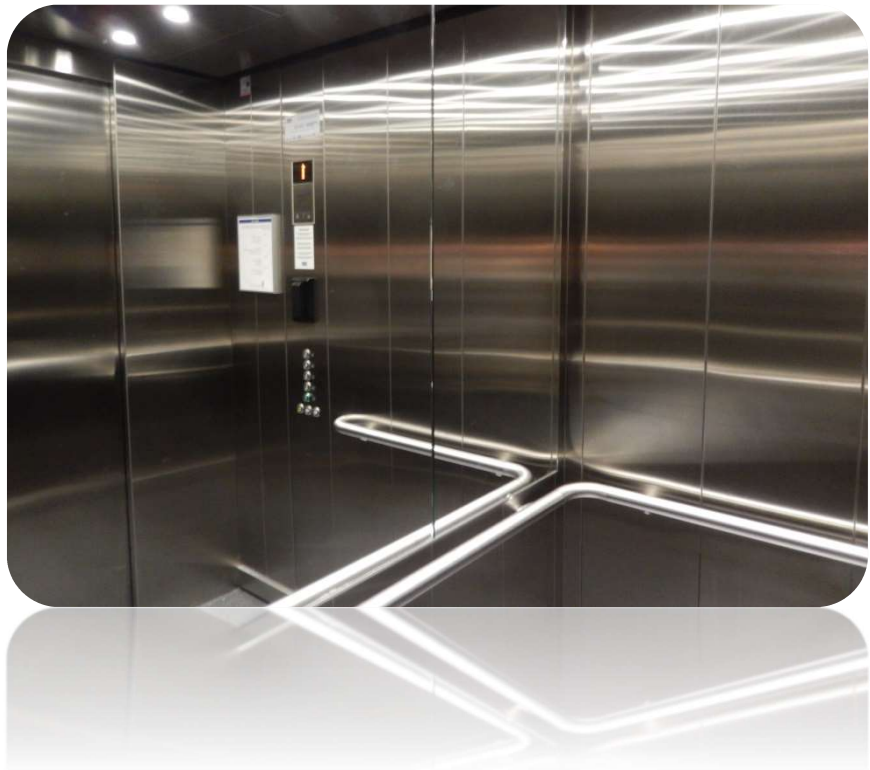

**June 30th, 2014**
**Planning and implementation time period:** March – April, 2014

**Research supervisor:** Pertti Saariluoma
**Researcher:** Rebekah Rousi
**Research assistant:** Merja Lehtiharju

# Contents

# Introduction

Elevator usage is an entirely embodied transaction, in which a user (traveller), requests the elevator traditionally by pressing the elevator hall call button (recently also by alternative methods such as apps, and foyer-based destination controls), enters, and controls the elevator's destination by pushing a numbered button. Inside the elevator cab, a human user, their body and psyche are subject to numerous factors, both physical and social. These qualities of elevator usage are often brought to the fore by the media, either in television shows or movies, where they are used to demonstrate various social dynamics – tension, love, anger, reluctance, empathy, and assistance to name a few. They are also utilized to demonstrate and emphasise fear, through exaggerating the potential dangers of their technical physics (height, speed, enclosure, suspension and unreliability) and exposing the vulnerability of the body within the cab space.

At any one time humans experience phenomena and the world around them both consciously and unconsciously. The reason being, is that although the body is constantly receiving and perceiving information from its surrounding environment, because of limitations in cognitive load capacity and attention, and the importance of prioritisation, people cannot consciously represent or be aware of every detail of what they encounter (Berridge & Winkielman 2003; Chalmers 1990). Conscious experience occurs when the normality of systems, processes and other phenomena is interrupted (Chalmers 2004; Peirce 2009). Thus, while a great part of information received through our body's sensory system remains unrepresented – or not consciously experienced – irregular occurrences and phenomena which are sensed (perceived) to be not normal, becomes represented or experienced (Berridge & Winkielman; Chalmers 1990; Peirce 2009). This is one of the body's evolutionary survival mechanisms, in order to prepare the individual for necessary action (Frijda 1988; Hekkert 2006).

There is an intricate relationship between bodily feelings or sensations and our subsequent emotions. In human-technology interaction (HTI), the much cited appraisal theory (Frijda 1988; Ortony & Turner 1990), explains how our emotions, and even higher order thinking (top-down processes), are subject to the way humans evaluate phenomena in terms of their own well-being. These appraisals are based on a mixture of both represented and unrepresented information – the primal reactions due to immediate evaluation of the benefit or threat perceived; and the emotions derived from higher cognitive processing, whereby information is evaluated against further criteria instilled through processes of learning and acculturation (Brave & Nass 2007; Jordan 2000).

Currently there is much discussion regarding the biological, social and psychological differences between men and women, which have developed through both evolutionary and cultural processes. One major finding in the field of neuroscience has been that of the differences in mirror neuron activity between men and women. Mirror neuron activity, shown through *mu* suppressions, or suppressions in the ventral premotor cortex has been connected with the brain mirroring the physical and emotional activities of other people (Gaag, Minderaa & Keysers 2007). In other words, for example when we see someone get hurt, our brain reacts in a way that it would if we ourselves get hurt. Which would suggest a biological basis for empathy, and given that women are found to have increased mirror neuron activity, one possible effect would be that women possess higher levels of empathy than men. Thus, if women have higher empathy levels than men, or are able to neurologically mirror the emotions and physical effects of people being observed, would this mean that women could possibly be used as instruments to observe and report other people's emotions? Say, in HTI circumstances?

This study set out to examine the relationship between expressed emotions (i.e. that what people say they are feeling – i.e. sad, angry, happy etc.) and physical sensations – the connection between emotion and bodily experience. It additionally provided the opportunity to investigate how the neurological findings of gender differences can be observed in practice – what difference does it make in behaviour and judgment that we have varying levels of mirror neuron activity? The following report documents the study, procedure, results and findings.

## Research questions

Based on the issues noted in the introduction, three central research questions were formulated. These questions guided the research design and subsequent inquiry:

1. What are the bodily sensations of travelling in an old, relatively unreliable elevator as compared to a newly renovated elevator?

2. What is the relationship between expressed emotions and bodily sensations?

3. Does gender make a difference? If women possess a stronger ability to empathise, can they be used as a measure for gauging the bodily experience of others?

# The Study

This study was designed to examine the relationship between the bodily sensations of elevator user experience and expressed emotions. Thus, the semiotic process and physio-semantic relationship between articulated emotions (the emotions people report as experiencing) and what is felt within the body. For this purpose, two elevators known as contrasting in their physical and operational characteristics were chosen for the study. Both were KONE elevators, housed in Agora (five storey's high). One of the elevators was relatively new, and had been installed in 2013, and the other was a former goods-come-passenger elevator installed during the building's completion in 2000. The elevators can be seen in the images below.

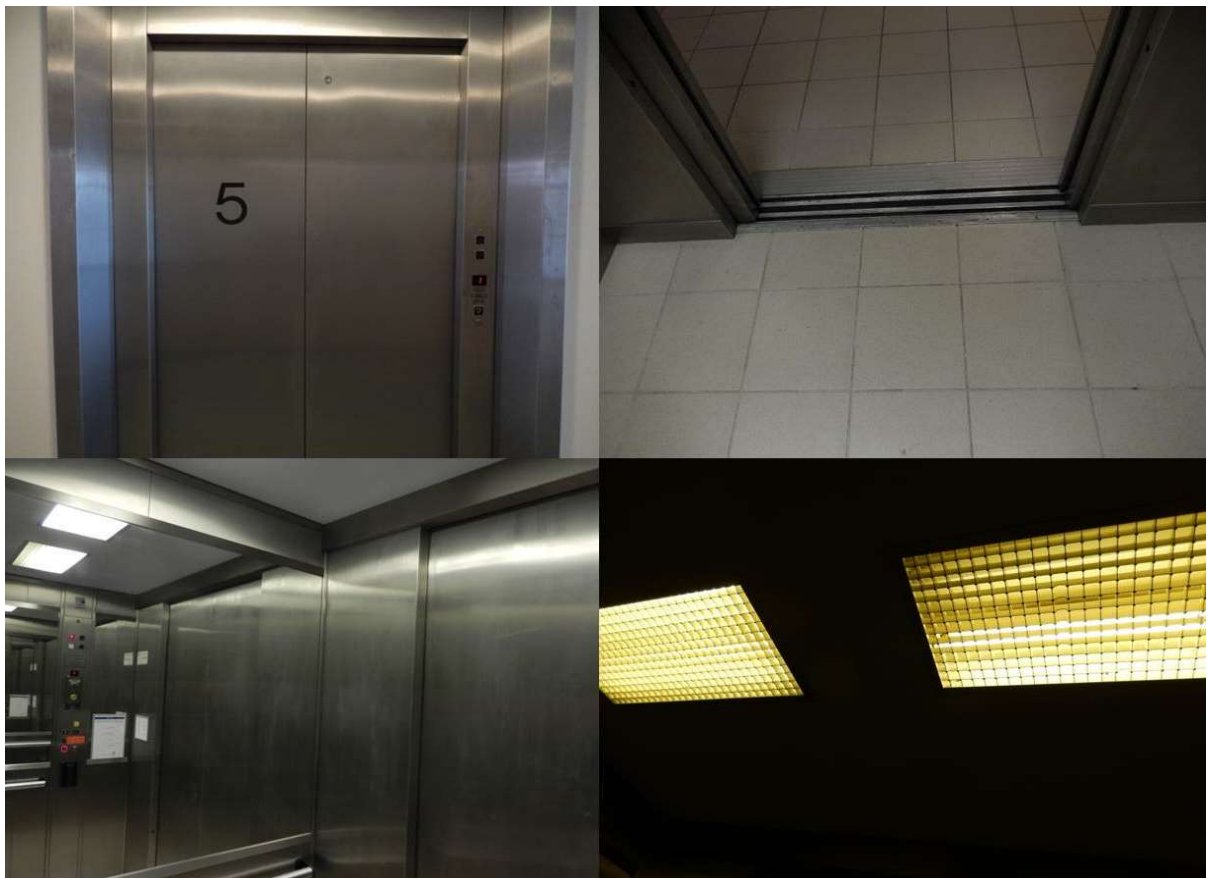

**Fig. 1: Old elevator – outer doors on fifth floor (upper left); floor and door opening (upper right); interior (lower left); ceiling lights (lower right)**

In figure 1 the old elevator can be seen, firstly in terms of its outward appearance from the hall call area. The image was taken on the fifth floor, however each floor is identical (a part from the numbers on the doors). This uniformity can be seen regard the continuation of the flooring from the main interior architectural design to the elevator flooring. The elevator cab aligns with the floors it arrives at, yet the travel is slow and the cab tends to 'bounce' and move with inconsistency. In the lower left-hand picture it can be seen that the cab is lined with clinical-like stainless steel, and the button of one floor is covered with orange tape to prevent pressing – this floor is currently undergoing renovations. Ceiling lights are industrial fluorescents, giving a bright, however grey impression.

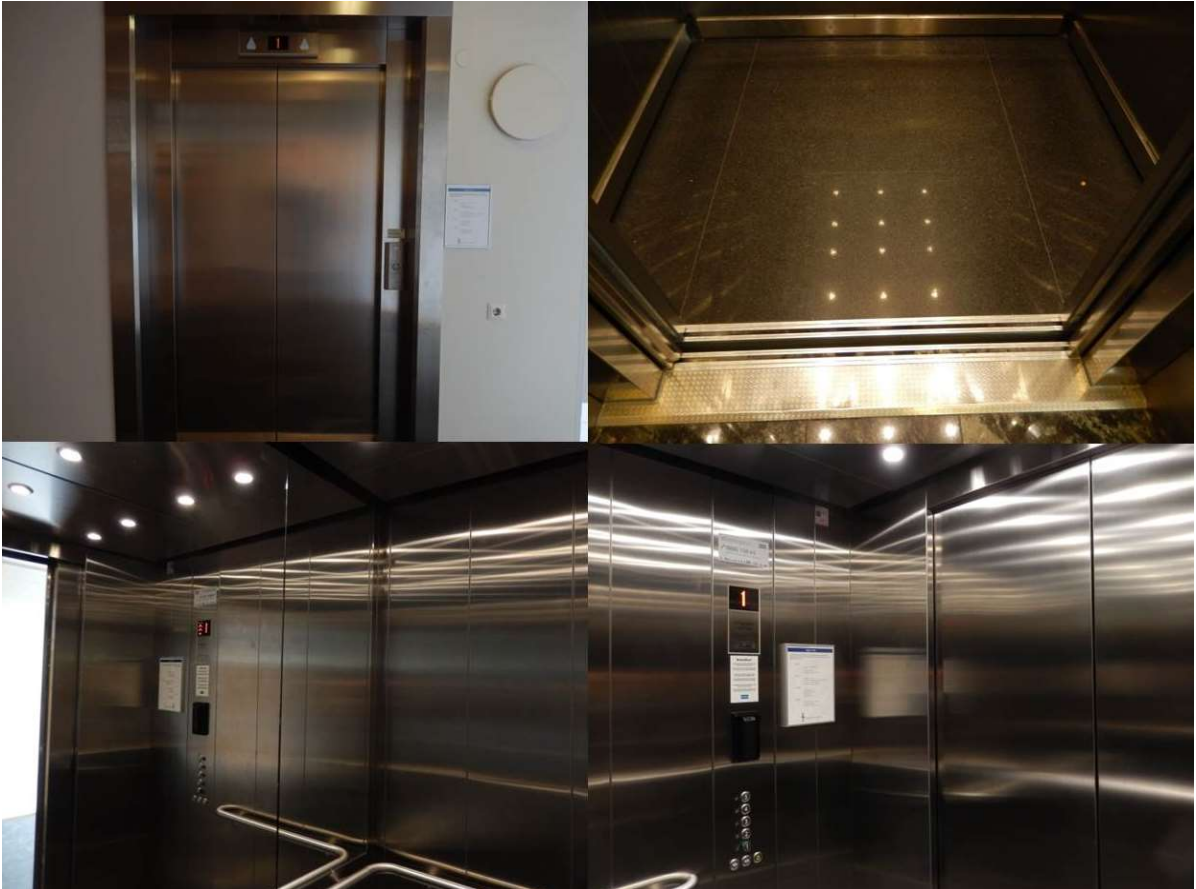

**Fig. 2: New elevator – outer doors on first floor (upper left); floor and door opening (upper right); interior through mirror (lower left); interior control corner (lower right)**

The new elevator, as seen in figure 2 indicates the floor on which the elevator is at via a display located at the top of the doors, as opposed to the old elevator which displays this information directly above the hall call button. The new elevator features a female voice informing the passengers of the current location of the elevator. Similarly to the old elevator, this elevator is also lined with stainless steel panels and features half wall mirrors on either side of the cab. The hand railing is curved in the new elevator, whereas it ends in sharper corners in the old elevator. Further, the controls take an upside-down T form, with the door controls and emergency button alone the bottom row and the floor buttons in a logical sequence going up. Whereas, the old elevator presents the controls in a rectangular order, with the first floor button at the bottom and emergency button at the top.

There are several aspects that make this elevator more appealing in comparison to the old elevator. Firstly, the stainless steel is highly polished and presented in thinner vertical panels. As seen in the images above, the combination of these factors with the grid of small LEDs generates a 'light wave' effect – a net or grid of light waves reflected off the panels generating a diffused, ambient lighting effect. Furthermore, careful attention has been placed on the flooring. Rather than trying to directly match the dark marbled granite flooring of the foyer, steps have been taken to select a granite-like flooring which comprises highly reflective greys and silvers, which adds depth to the experience, in addition to light.

The study was conducted by participants in pairs or groups of three, whereby one participant (the control, Part A), would be told to fill out a questionnaire based on their own perspective of the elevator experiences and the remaining participants (Part B) were told separately to fill in the questionnaire from the perspective of Part A. The idea was to see how well the answers provided by Part B matched those of Part A (taking into account gender).

Experiments took place during March-April, 2014. It took on average 20 minutes to conduct the experiment. Participants were required to travel in each elevator from the top floor (fifth storey) to the ground floor and up again. Minimal discussion was encouraged, and participants were given the instructions to *feel* the sensations in their body – and on behalf of Part A. Upon exiting the elevators, participants were instructed to move to separate locations to fill in the questionnaires. Participants could ask for help if they needed further clarification regarding the terminology. Participants were recruited via university email lists and word-of-mouth. They were rewarded with a small chocolate gift.

## Questionnaire design

The questionnaire comprised four sections: 1) background information – gender, year of birth, nationality and professional/study field; 2) one emotional construct (word describing the emotion); 3) the indication of where they feel the bodily sensations, by colouring a human silhouette (yellow for the 'hot spots' where they felt the sensation and blue for the 'cold spots' which were completely unaffected); and 4) semantic differential – a) bodily feeling inside the elevator; b) elevator design; and c) social emotion.

### Emotion and bodily sensations

The technique of asking participants to identify which areas of the body experienced feelings or sensations during elevator travel was influenced by a study on bodily emotions by Nummenmaa, Glerean, Hari and Hietanen (2013). Nummenmaa et al.'s study concentrated on identifying the areas in the body which were connected to basic and non-basic emotional states reported by participants in response to observing affective stimuli. The emotions focused on in Nummenmaa et al.'s (2013) study were: basic - anger, fear, disgust, happiness, sadness, surprise, and neutral; non-basic – anxiety, love, depression, contempt, pride, shame and envy. In this study we wanted to further this examination by applying the same technique to isolate bodily sensations in relation to the entirely embodied experience of elevator travel, in connection to qualitative emotional constructs provided in one word. After writing an experienced emotion (construct) down in the space provided, participants were asked to colour in a silhouette – blue for the cold unaffected regions and yellow for the regions in which they experienced some kind of sensation.

### Semantic differential categories

Semantic differential is a method developed by Osgood, Suci and Tannenbaum (1957) designed to test the relationship between qualitative constructs – often adjectives – and syntactic properties generally found in product design. There are numerous forms and types of semantic differential and visual syntactic scales (Warell 2004), a common form being a Likert-scale like questionnaire posing sets of binary polar adjectives (e.g. good-bad). Where in many opinion rating scales numbers represent a relationship with negative-positive valence, in semantic differential scales the positive-negative valence is dependent on the construct's relationship to the evaluation subject and context, i.e., the term "traditional" may be desired in certain contexts in connection to certain objects or properties, in others it may mean datedness, and lagging behind the times.

The main idea is to *test* the meanings of the words in connection to chosen subjects. The categories were formulated according to three dimensions of elevator experience: what is felt in the body; how the elevator's design portrays an experience to the user; and what is felt socially. Based on these categories, bipolar adjective pair groupings were formulated: 1) bodily feeling – space, movement consistency, security and safety; 2) elevator – space, aesthetic, emotional design, and safety; and 3) social emotions – interactivity, affectivity and self-reflexivity. The following tables detail the bipolar adjective pairs according to their groupings (tables 1-3).

*TABLE 1: Bodily feeling inside the elevator*

| CONSTRUCT DIMENSIONS | Negative constructs | Positive constructs |
|---|---|---|
| **Space** | crowded | roomy |
|  | distanced | intimate |
| **Movement consistency** | bumpy | smooth |
|  | swaying | firm |
|  | butterflies | neutral |
| **Security and safety** | vulnerable | protected |
|  | comfortable | uncomfortable |
|  | unstable | stable |
|  | dipping | steady |
|  | tingling | strong |

*TABLE 2: Elevator design*

| CONSTRUCT DIMENSIONS | Negative constructs | Positive constructs |
|---|---|---|
| **Space** | claustrophobic | spacious |
| **Aesthetic** | ugly | beautiful |
|  | dull | fun |
|  | boring | interesting |
|  | dated | timeless |
| **Emotional design** | unconvincing | reassuring |
|  | stressful | calming |
| **Safety** | unsafe | safe |

*TABLE 3: Social emotions*

| CONSTRUCT DIMENSIONS | Negative constructs | Positive constructs |
|---|---|---|
| **Connectivity** | isolated | connected |
| **Interactivity** | uninviting | inviting |
|  | shy | outgoing |
|  | anti-social | social |
| **Affectivity** | anxious | indifferent |
|  | uneasy | easy |
|  | irritating | relaxing |
| **Self-reflexivity** | self-conscious | confident |
|  | embarrassed*** | proud***[1] |

The adjectives were chosen from the data of previous elevator studies, in which people described their feelings towards and inside of elevators. Some of the constructs are abstract such as "tingling", and "butterflies" as they attempted to describe the 'difficult to describe' sensations experienced during movement and interaction which are often induced by unexpected occurrences.

---

[1] *** please note that this construct pair was only given to Part. A's in order to observe whether or not they somehow sensed being observed.

## Participants

The study was planned and implemented during March and April, 2014. Participants were sought through university email lists and word-of-mouth. In total, 45 participants were recruited, 23 females and 22 males. Participants took part in the study in pairs and groups (max. 3). For every group one participant was a Part A (basic participant) and the rest were Part B (those who guessed Part A's experiences). The study was implemented 20 times, meaning that there were 20 Part As and 25 Part Bs. Of the Part As, 9 were female and 11 were male. Of the Part Bs 11 were male and 14 were female. Participants ranged in age from 24 to 64 years (average 34,7), and were either students or staff of the university in the fields of information systems, information technology, cognitive science and music psychology.

# Results

The results of this experiment are multi-faceted, and comprise numerous sections. This results chapter presents the findings in the following sections: emotional constructs; bodily sensations and semantic differential.

## Emotional constructs

In Step 2 (after the background details) of the questionnaire participants were requested to provide one word to describe their emotions before colouring in a human silhouette to identify the location of the bodily sensation they experienced in the elevator. In the analysis the constructs grouped according to responses by Part As and Part Bs, and then they were subsequently divided into categories according to common characteristics observed in each instance. Table 4 outlines the raw constructs provided by Parts A and B in response to experiencing the old elevator.

*TABLE 4: emotional constructs provided in response to the old elevator*

| GROUPINGS | PART A | GROUPINGS | PART B |
|---|---|---|---|
| **Positively aroused** | Funny | **Positively aroused** | Curious |
| | Anticipation | | Fun |
| | Amused | | Happy |
| | Exciting | | Amused |
| **Negatively aroused** | Nervous | **Negatively aroused** | Alert |
| | Uneasy | | Uneasy |
| | Scary | | Nervous |
| | Shy | | Tensed |
| **Negative temporal** | Impatience | | Suspicion |
| | Slow | | Uncomfortable |
| | Boring | | Irritated |
| | Waiting | **Negative temporal** | Bored |
| | Restless | | Impatient |
| **Elevator characteristics** | Restricted | | Busy |
| | Bumpy | | Waiting |
| | Uneven | **Positively passive** | OK |
| | Grey | | Relaxed |
| **Neutrality** | Neutralness | **Neutrality** | Calm |
| | Normal | | Indifferent |
| **Mood** | Thoughtful | | |

These were sorted into the following groups: positively aroused, negatively aroused, negative temporal (time and waiting related), elevator characteristics, neutrality, mood and positively passive (calm, relaxed).

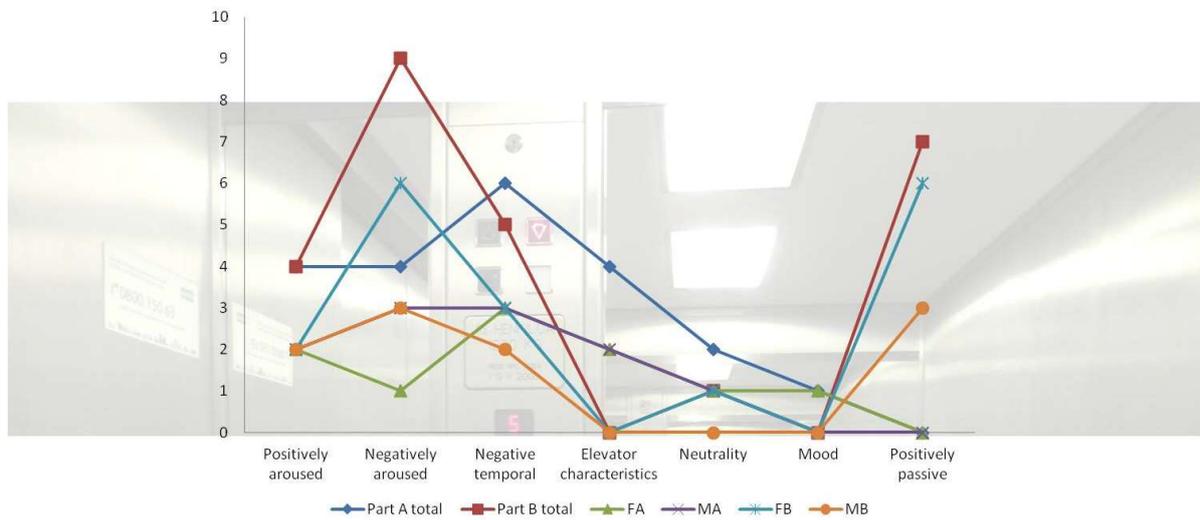
**Fig. 3 emotional constructs of the old elevator**

The diagram above shows that Part As supplied a total of: 4 positively aroused and 4 negatively aroused constructs; 6 negative temporal constructs; 4 constructs pertaining elevator characteristics; 2 related to neutrality; and 1 relating to mood. This is contrasted with constructs provided by Part Bs, who provided: 4 positively aroused constructs; **9 negatively aroused**; 5 negative temporal; none relating to elevator characteristics or mood; and **7 related to positively passive**.

When analysing the constructs by gender it is noticed that Female Part As provided: 2 positively aroused; 1 negatively aroused; **3 negative temporal**; 2 about the elevator characteristics; 1 neutrality construct; and 1 about mood. Male Part As provided very similar constructs: 2 positively aroused; **3 negatively aroused**; **3 negative temporal**; 2 about the elevator characteristics; 1 relating to neutrality. In other words, while both men and women provided negative constructs relating to time and waiting, the men in particular had noted negative arousal. However, for the most part sentiments shared between the genders were quite similar.

In relation to Part Bs, female Part Bs provided: 2 positively aroused; **6 negatively aroused**; 3 negative temporal; no elevator characteristics; 1 about neutrality; none about mood; and **6 positively passive**. Male Part Bs gave: 2 positively aroused; **3 negatively aroused**; 2 negative temporal; none regarding elevator characteristics, neutrality or mood; and **3 positively passive**. While as reflected in the results, the Male Part B group was slightly smaller than the female one, there are still similarities between what the female and male Part Bs provide.

What is interesting to note is that Part Bs were more negative, in relation to the affect (negative arousal) than Part As, yet Part As seemed to feel the time factor more negatively. Female Part Bs varied quite substantially from Female As in their experience of negative arousal. Part As did not remark on the positive passivity at all as compared to Part Bs.

*TABLE 5: emotional constructs provided in response to the new elevator*

| GROUPINGS | PART A | GROUPINGS | PART B |
|---|---|---|---|
| **Positively aroused** | surprised | **Positively aroused** | Emancipated |
| | amused | | Mild delight |
| | interesting | | Happy |
| | prank-like | | Good |
| **Negatively aroused** | anxiety | | Curious |
| **Negative temporal** | restless | | Interested |
| | impatience | **Negatively aroused** | Not relaxed |
| **Elevator characteristics** | techy | | Tensed |
| **Positively passive** | Relaxation | | Awkwardness |
| | pleasant | | Uncomfortable |
| | flowing | | Uneasy |
| **Neutrality** | normal | | Mild tension |
| | neutral | | Distracted |
| **Safety Reliability** | safe | **Positively passive** | Calm |
| | stable | | Cool |
| | | | Relaxed |
| | | | Easy |
| | | | Comfortable |
| | | **Neutrality** | Neutral |

These emotional constructs were sorted into the following categories: positively aroused, negatively aroused, negative temporal, elevator characteristics, positively passive, neutrality, and safety reliability. The reason that the categories of these two elevators differ is due to the differences and prominences of constructs supplied by the participants.

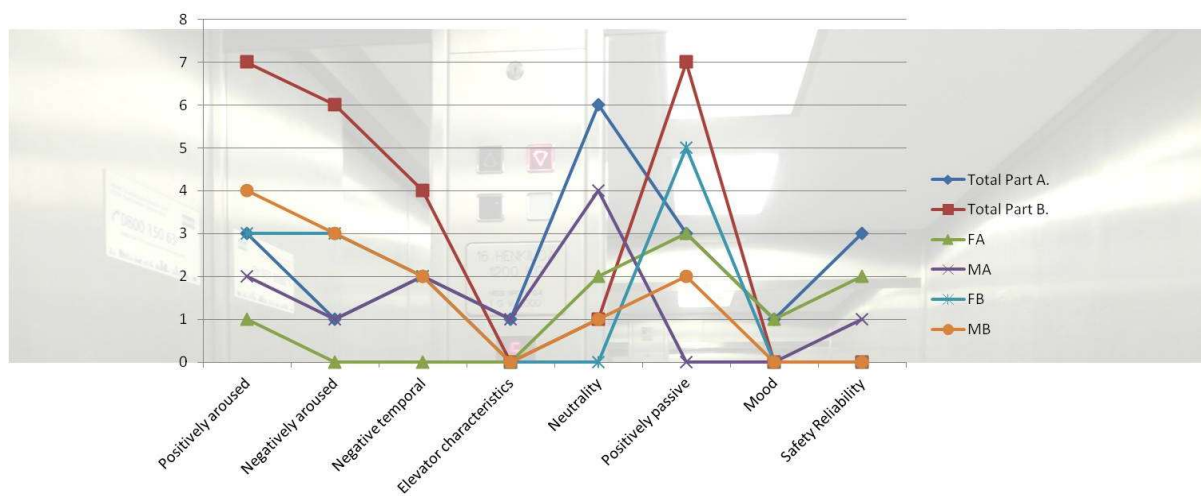

**Fig. 4 Emotional constructs of the new elevator**

The diagram above shows that Part As supplied a total of: 3 positively aroused and 1 negatively aroused constructs; 2 negative temporal constructs; 1 construct pertaining elevator characteristics; **6 related to neutrality**; 3 positively passive constructs; 1 relating to mood; and 3 referring to safety and reliability. This is contrasted with constructs provided by Part Bs, who provided: **7 positively aroused constructs**; **6 negatively aroused**; 4 referring to negative temporal; none relating to elevator characteristics or mood; **7 related to positively passive**; and none concerning safety or reliability.

When analysing the constructs by gender it is noticed that Female Part As provided: 1 positively aroused; no negatively aroused; none regarding negative temporal; none about the elevator characteristics; 2 neutrality constructs; **3 relating to positively passive**; 1 about mood and 2 about safety and reliability. Male Part As provided: 2 positively aroused; 1 negatively aroused; 2 negative temporal; 1 about the elevator characteristics; **4 relating to neutrality**; no positively passive or mood constructs; and 1 about safety and reliability. Here it is noticed that there is less attention drawn towards the specific elevator characteristics and more towards notions of neutrality and positive passive experience – i.e., the ability to relax, enjoy and not notice the elevator travel.

In relation to Part Bs, female Part Bs provided: 3 positively aroused; 3 negatively aroused; 2 negative temporal; none about elevator characteristics nor neutrality; none about mood; **5 positively passive**; and none about safety and reliability. Male Part Bs gave: **4 positively aroused**; 3 negatively aroused; 2 negative temporal; none regarding elevator characteristics; 1 about; 2 positively passive; and none about mood or safety and reliability. The results were quite even in regards to positively and negatively aroused, negative temporal and the rest. The only larger difference can be seen in relation to positively passive constructs – females placed greater emphasis on relaxation and calmness. Yet, neither female or male Part Bs referred to the positive experience of safety and reliability as the Part As had.

Part Bs in general, gave more positively aroused constructs than Part As, but this was combined with an almost equal amount of negatively aroused constructs from Part Bs. What was mostly expressed amongst Part As were sentiments of neutrality and feeling normal in the efficient new elevator. Part Bs enhanced this notion with another strong emphasis on positive passivity (calm and relaxed). The female Part Bs expressed the most amount of constructs regarding positive passivity, while the male Part Bs provided the most constructs regarding positive arousal. Overall there seemed to be an *amplification* in what Part Bs mentioned in regards to how *they* felt Part As experienced the elevators as compared to how Part Bs actually felt.

**Bodily sensations**

Upon searching for methods to measure the relationship between expressed emotions and bodily sensations, the Nummenmaa et al. (2013) body emotions study was discovered. As explained above, participants were asked to identify, by colouring a silhouette, where they most felt sensations in their body (yellow) and where they least felt these sensations (blue), and likewise Part Bs were asked to colour in the body according to where they *felt* (i.e. through their imagination or through their own bodily sensations) the other person was feeling the sensations. The below figures are compilations of all of the responses for firstly the old elevator and then the new elevator.

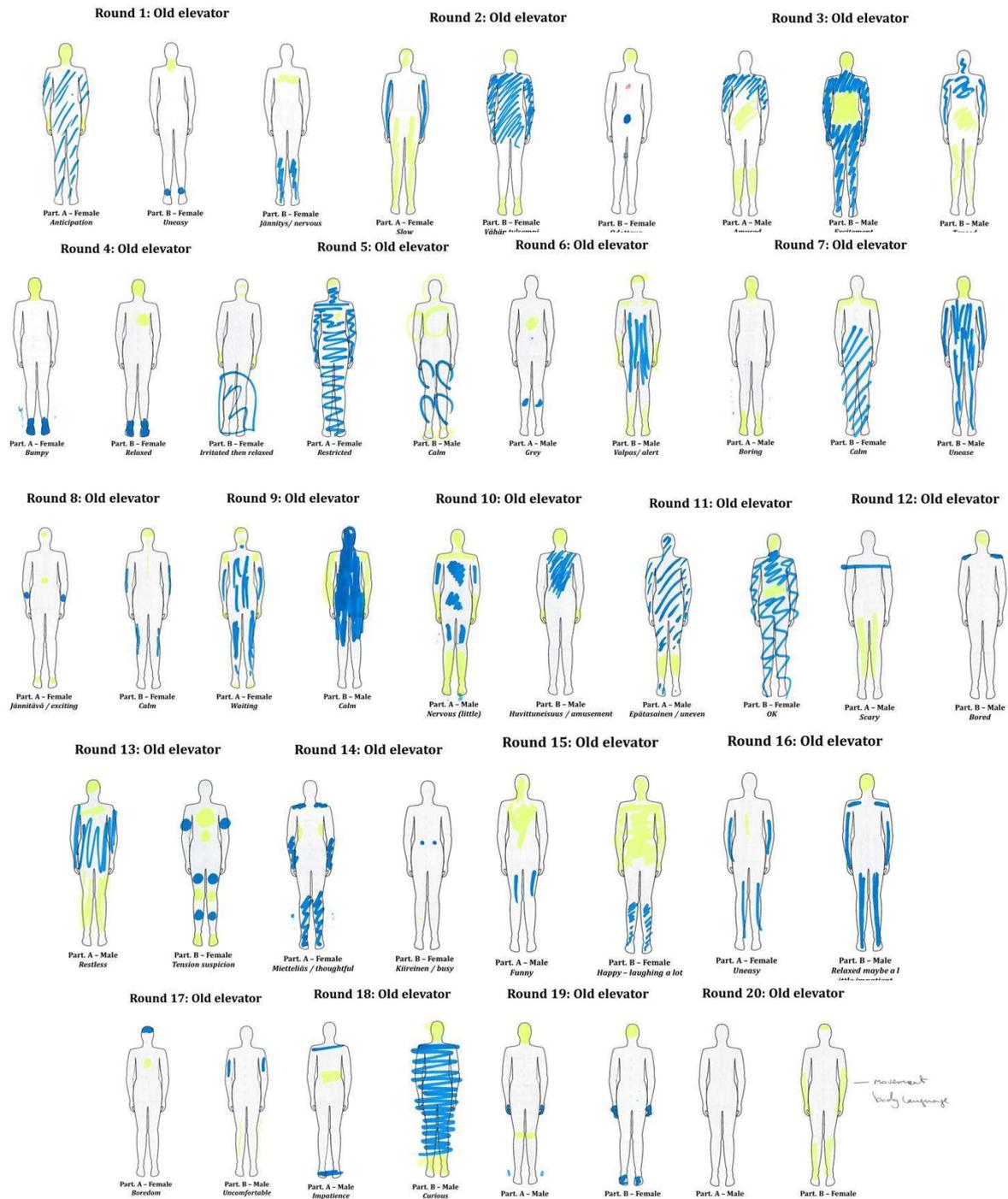

**Fig. 5: Bodily sensation silhouettes regarding the old elevator**

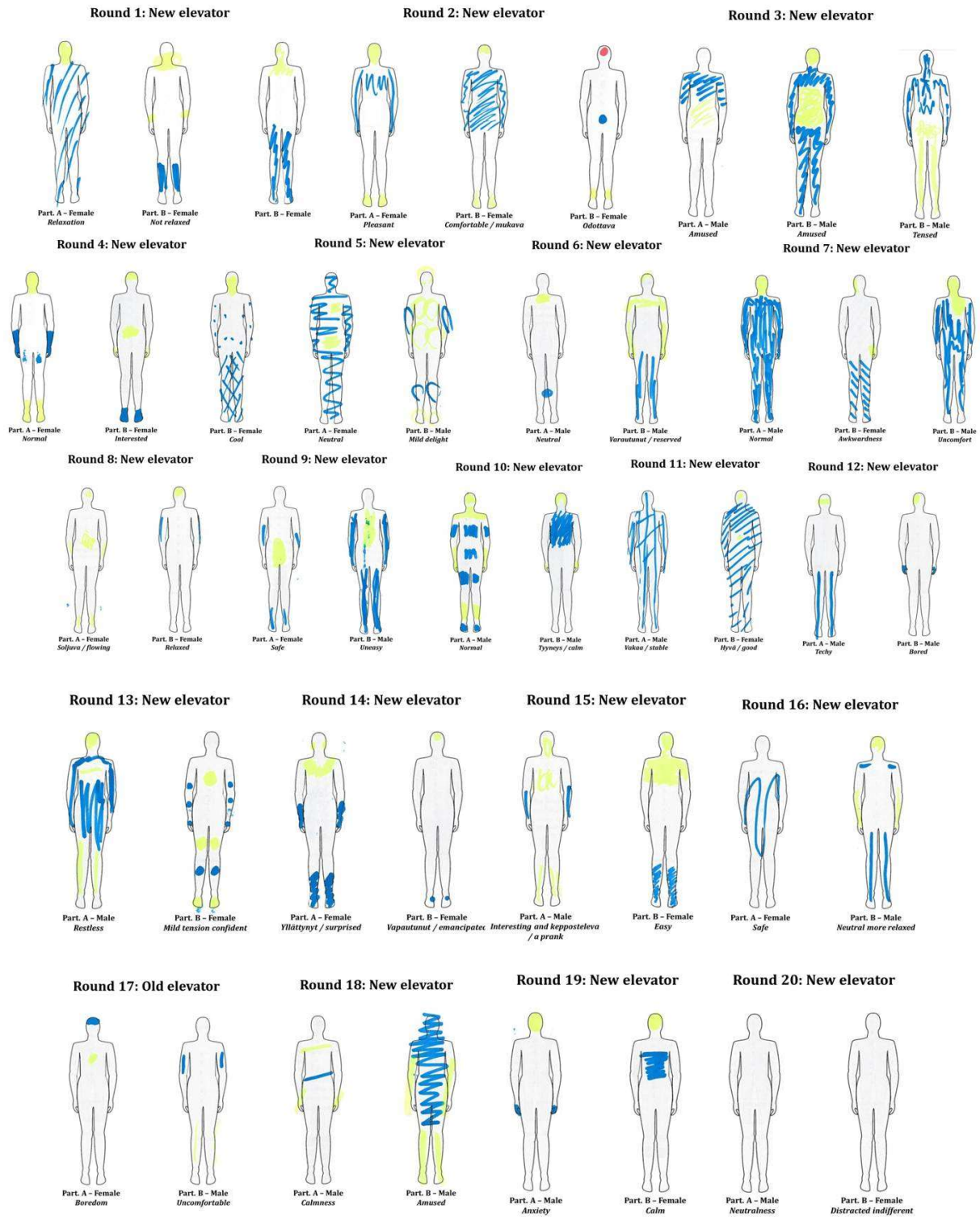

**Fig. 6: Bodily sensation silhouettes regarding the old elevator**

The silhouettes have been analysed according to the parts of the body in which participants have indicated their sensations. These body parts are: heads, shoulders, arms, hands, chests, stomach, pelvis, thighs, shins and feet. Results of the analysis can be seen in the tables below.

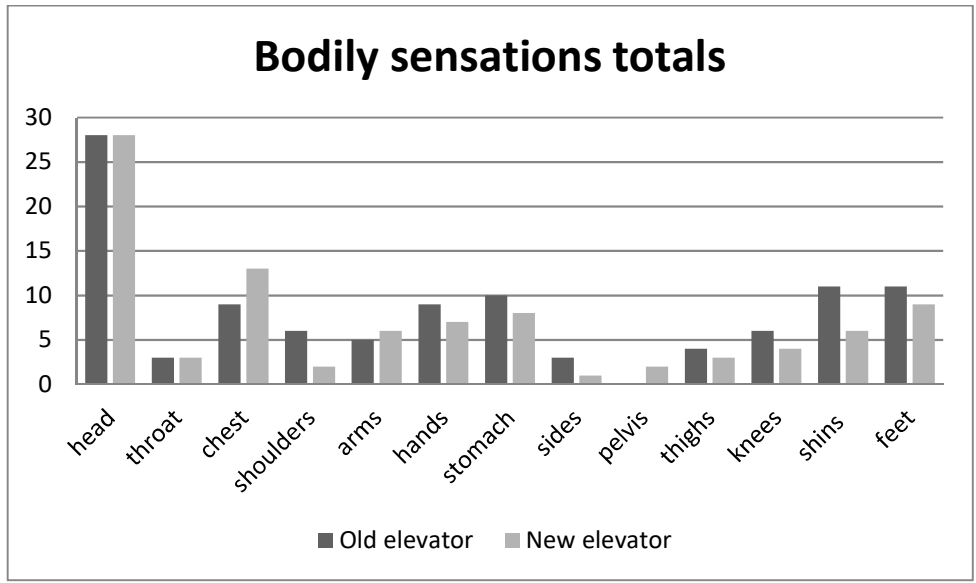

**Fig. 7: Bodily sensations totals – identified body parts per elevator**

The figure above shows that the area in which people experienced the most bodily sensations was in the head. This figure was equal (28) for both elevators. The next affected area was the chest region which was indicated more often in relation to the new elevator (13) than the old (9). The shins and feet were indicated 11 times in the case of the old elevator, while in the case of the new elevator the feet were indicated 9 times and the shins 6 times. Bodily sensations in the stomach were also experienced more often in the old elevator (10) than in the new (8). More bodily sensations were experienced in the hands in the old elevator (9) than in the new (7). Yet slightly more sensations were felt in the arms in the new elevator (6) than in the old (5). Sensations were felt more in the shoulders in the old elevator (6) than the new (2). Likewise, more sensations were felt in the knees in the old (6) than the new (4). There were slightly more sensations experienced in the thighs in the old elevator (4) than the new (3). Additionally, more people indicated that they felt, or believed Part A felt, sensations in their sides in the old elevator (3) than in the new (1). Three people indicated feeling sensations in their throats in both elevators. Finally, two indicated sensations in their pelvis in the new elevator (2).

The next graph shows the distribution of bodily sensations according to the experiment groups. As seen in the diagram, some groups indicated sensations, or believed sensations of Part Bs observing Part As, more than others. But this too is an interesting point of discussion.

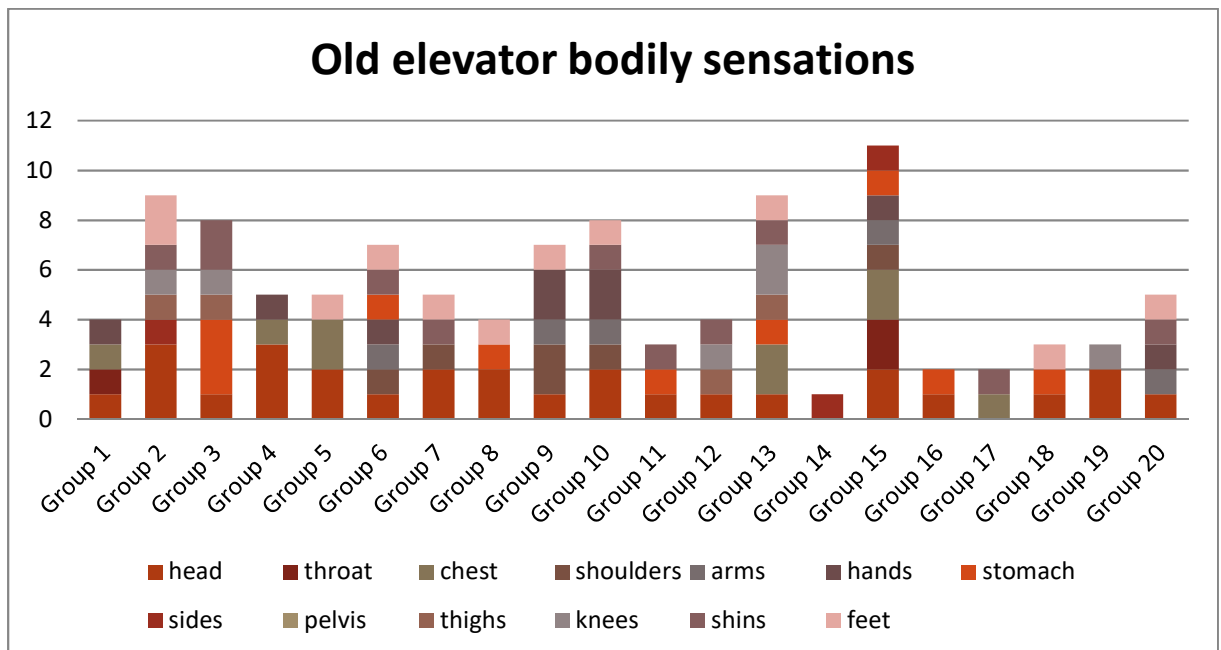

**Fig. 8: Identified bodily sensation areas per group – old elevator**

In the diagram it can be seen that in the case of group 1 (all females), each of the participants (Parts A and B) identified different areas in the body in which they felt sensations. The identified areas were concentrated towards the top half of the body. In group 2 (all females), **all of the participants (3) identified sensations in their head**, one in their throat (Part B), another (Part A) thighs, one knees, shins and the **two in their feet (Part A and B)**. Group 3 comprised all males and in this group **all the participants (3) indicated their stomach**, one head, one their thighs and knees and **two (Part A and B) shins**. In group 4 (all female) **all participants (3) identified their heads** one chest, and one hands. In group 5 (female Part A and male Part B) **both participants (2) indicated sensations in their heads and chests** and one (Part B) identified their feet. In group 6 the pair (both males) identified almost the opposite of one another. Part A identified the stomach, while Part B identified the head, shoulders, hands, shins and feet. In group 7 (2 males and 1 female) **two identified their head (Part A and male B)**, one (female B) identified shoulders, and another (Part A) shins and feet. Both participants identified their heads in group 8, then one stomach (Part A) and one feet (Part A). In group 9 **both participants (female and male) identified their shoulders and hands**, while one (Part A) identified their head and the other (Part B) identified their feet. In group 10 (both male) **both participants identified their heads**, one (Part A) in their shoulders and arms, **both in their hands**, and only Part A in their shins and feet.

In group 11 (Part A male, Part B female) there were no similarities, Part A felt sensations in his shins and Part B felt them in her head and stomach. Group 12 showed that Part A (male) felt the sensations in his thighs, knees and feet and Part B (male) felt it in his head. In group 13 (Part A male, Part B female) **both participants felt sensations in their chests and knees**. Part A felt them in his head, thighs and shins and Part B felt them in her stomach and feet. Group 14 (both female) is interesting in that both Parts identified their sides, yet Part A identified them as a site of sensation and Part B identified them as 'cold spots' (unaffected). In group 15 (Part A male, Part B female), **both Parts identified their heads, throats, chests and stomachs**, while Part B also identified her shoulders, arms and sides. In group 16 (Part A female, Part B male) Part A identified her stomach and Part B identified his head. In group 17 (Part A female, Part B male) Part A identified her chest and Part B identified his shins. Similarly to group 14, group 18 (two males) is also interesting in that they identified the opposites to each other. Part A identified his stomach as feeling the sensation, and the cold spots as being the shoulders and feet, while Part B identified the cold spots as everywhere else but head and feet. In group 19 (Part A male, Part B female) **both Parts felt sensations in their heads** and Part A also felt sensations in his knees. Group 20 (male Part A and female Part B) was an outstanding case, whereby Part B explained that she had only marked sensations in the head, arms, hands, shins and thighs due to the movement inside the elevator. Otherwise, she did not *feel* that Part A was experiencing any particular emotions or sensations. Part A did not mark any areas as he told that he *did not* feel anything.

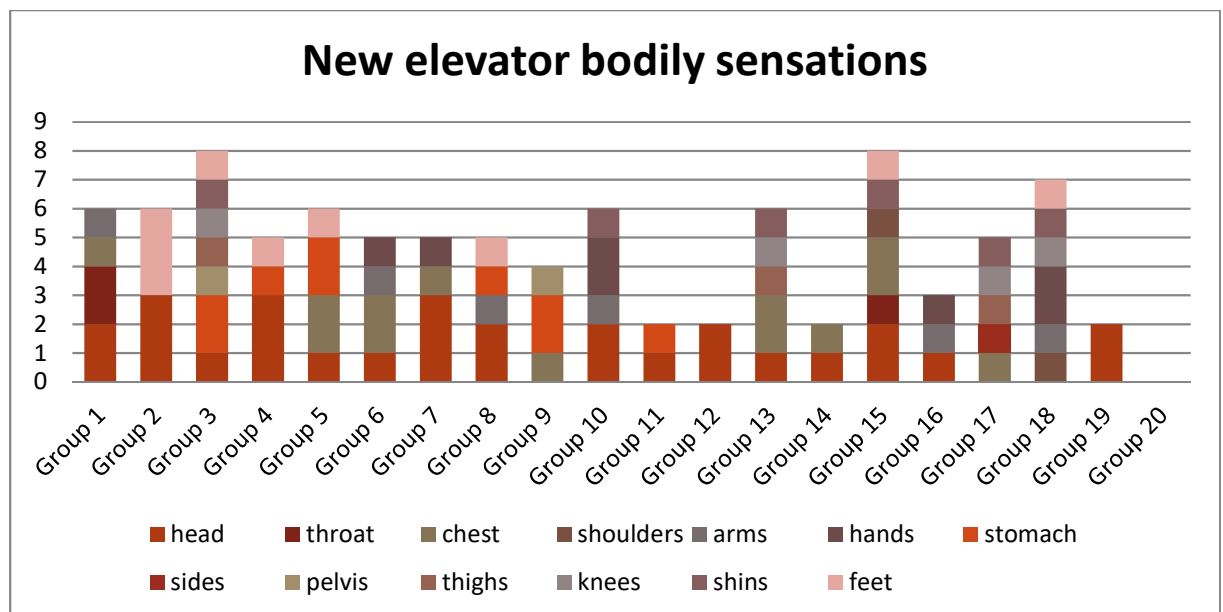

Fig. 9. Identified bodily sensation areas per group – new elevator

In group 1 (all female), **all participants identified sensations in their throat, two (Part A and Part B) identified the head**, and one (the other Part B) identified the hands. In group 2 (all female), **all participants identified sensations in their heads and feet**. In group 3 (all male), **Part A and one Part B identified sensations in their stomachs**, one Part B their head, chest and sides, the other Part B their thighs, knees, shins and feet. In group 4 (all female) **all participants identified their heads**, Part A her feet and one Part B her stomach. In group 5 (Part A female, Part B male) **both participants identified their chests and stomachs**, Part B their head and feet. In group 6 (both male) **both participants identified their chests**, while Part B also identified his head, shoulders, arms and hands. In group 7 (Part A male and Parts B male and female), **all participants identified their heads**, one Part B hands and the other Part B (male) chest. In group 8 (Part A female, Part B male), **both participants identified their heads**, then Part A also identified her stomach, shins and feet. In group 9 (Part A female, Part B male), **both participants identified their stomachs**, Part A identified her pelvis and Part B identified his chest. In group 10 (both male), **both participants identified their heads and hands**, Part A also identified his throat, shoulders and shins.

In group 11 (Part A male, Part B female) only Part B identified sensations in her head and stomach. In group 12 (both male), **both participants identified sensations in their heads**. In group 13 (Part A male, Part B female), **both participants reported sensations in their chests and thighs**, Part A identified sensations in his head, knees and shins, and Part B identified sensations in her feet. In group 14 (both male), **both participants noted sensations in their heads**, while Part A also indicated sensations in his chest. In group 15 (Part A male, Part B female), **both participants identified sensations in their heads and**

**chests**, Part A identified their shins and feet, and Part B identified their throat, shoulders and arms. In group 16 (Part A female, Part B male), Part A did not report sensations, but Part B showed sensations in his head, arms and hands. In group 17 (Part A female, Part B male), Part A identified sensations in her chest and part B identified his thighs, knees and shins. In group 18 (both male), both participants identified sensations in their hands. Part A identified sensations in his chest and shoulders, and Part B identified his arms, shins and feet. In group 19 (Part A male, Part B female), **both participants identified their heads**. And once again, group 20 (Part A male, Part B female) was an interesting case, **where neither of the participants indicated any sensations**. Part B explained (in conversation separately with the researcher) that she was perplexed that she could not detect anything from the participant, but Part A had noted on the questionnaire that he had not experienced any sensations.

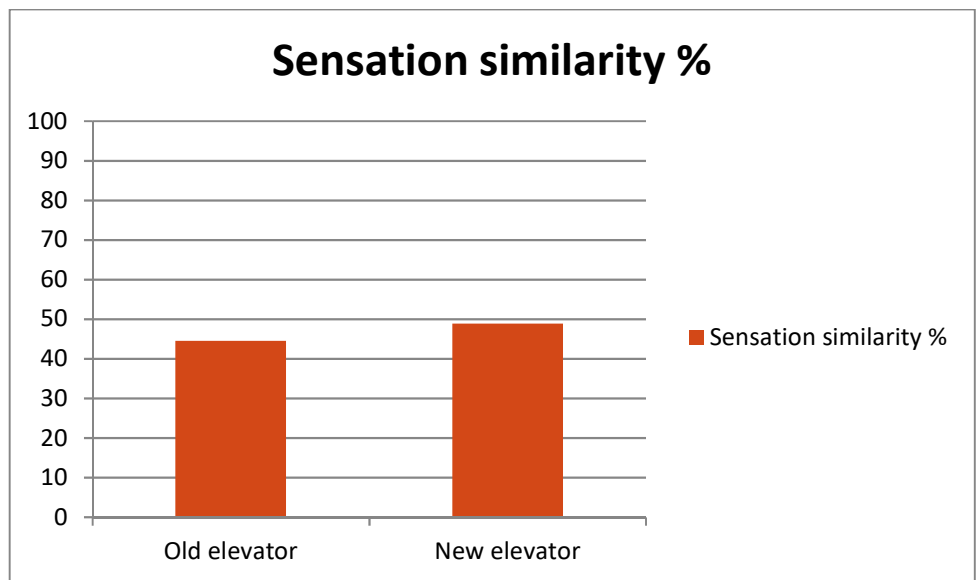

Fig. 10. Comparison of the rates of similarity between the elevators

The graph above shows the rates of similarity of identified bodily sensations between the elevators. It can be seen that in all cases, almost half of the time Parts B were able to accurately identify where in the body Part As were experiencing sensations. However, as also can be seen, identification in response to the new elevator was slightly more accurate. This may indicate that the more neutral, or stable the elevator experience, the more standardised the sensation in the body. This observation is supported by what can be seen in the tables below which match the emotional constructs to bodily experiences.

*TABLE 6: Old elevator emotional constructs with bodily sensations*

| BODY PARTS | Part A female constructs | Part A male constructs | Part B female constructs | Part B male constructs |
|---|---|---|---|---|
| **Head** | Anticipation<br>Slow<br>Bumpy<br>Restricted<br>Exciting<br>Waiting | Boring<br>Nervous<br>Restless<br>Funny<br>Shy | More boring<br>Waiting<br>Relaxed<br>Irritated then relaxed<br>Calm<br>OK<br>Happy<br>Fun | Excitement<br>Calm<br>Alert<br>Unease<br>Amusement<br>Bored<br>Relaxed a little impatient<br>Curious |
|           |              |             | Indifferent         |              |
| **Throat** | Bumpy       | Boring      | Anticipation        | Unease       |
|           |              | Nervous     | Happy               | Curious      |
|           |              | Funny       |                     |              |
|           |              | Shy         |                     |              |
| **Chest** | Restricted   | Grey        | Nervous             | Excitement   |
|           | Uneasy       | Restless    | Waiting             | Calm         |
|           | Boredom      | Funny       | Tension suspicion   |              |
|           |              |             | Relaxed             |              |
|           |              |             | Happy               |              |
| **Shoulders** | Waiting  |             | Calm                | Alert        |
|           |              |             | Happy               | Calm         |
| **Arms**  |              |             | Nervous             | Calm         |
|           |              |             | Happy               | Calm         |
|           |              |             | Indifferent         |              |
| **Hands** | Anticipation |             | Irritated then      | Alert        |
|           | Waiting      |             | relaxed             | Calm         |
|           |              |             | Happy               | Nervous      |
|           |              |             | Indifferent         | Amusement    |
| **Stomach** | Exciting   | Grey        | OK                  | Amused       |
|           |              | Funny       | Tension suspicion   | Excitement   |
|           |              | Impatience  | Happy               | Tensed       |
|           |              |             |                     | Calm         |
| **Sides** | Slow         |             | Happy               |              |
| **Pelvis** |             | Amused      |                     | Tensed       |
| **Thighs** | Slow        | Scary       |                     | Tensed       |
|           |              | Restless    |                     |              |
| **Knees** | Slow         | Scary       | Amused              | Tensed       |
|           |              | Restless    | Tension suspicion   |              |
|           |              | Shy         |                     |              |
| **Shins** | Slow         | Amused      | Scary               | Tensed       |
|           |              | Alert       | Indifferent         | Nervous      |
|           |              |             |                     | Uneven       |
|           |              |             |                     | Restless     |
|           |              |             |                     | Uncomfortable |
| **Feet**  | Slow         | Boring      | More boring         | Calm         |
|           | Exciting     | Nervous     | Tension suspicion   | Alert        |
|           | Waiting      |             |                     | Curious      |
|           |              |             |                     | Indifferent  |

It can be seen that the most common part of the body affected by the elevator experience was the head. The head was attached to both positive and negative emotional constructs, aroused and passive. Females indicated the head slightly more often than males and Part Bs identified the head area slight more in general. The throat was identified mostly by male Part As – three negatively aroused emotions plus one positively aroused. Part Bs (male and female) were equal in identifying sensations in their throats. All of the emotional constructs were aroused, both positive and negative. All the emotional constructs Part As identified as connected to the chest area were aroused except 'grey'. More **female Part Bs** identified the chest area than male Part Bs, and there was a mixture of emotional constructs from positive to negative, aroused to passive. The shoulders were more commonly identified amongst Part Bs. Again, there was a mixture of emotional constructs – two positively passive

(calm) and two aroused (positive and negative). Only one Part A (female) identified the shoulders in connection to the experience of 'waiting'. The arms were not identified at all by Part As. There was a mixture of emotions expressed by female Part Bs in connection with the arms, both cases in which **male Part Bs identified the arms they expressed the emotion 'calm'**.

**Only female Part As identified the hands**, and in both cases this related to an aspect of time and anxiety – anticipation and waiting. Both female and male Part Bs identified the hands, but there was a mixture of emotional constructs also present. **A common thread amongst the identification of the stomach area in relation to both Part As and Bs was that of excitement, impatience, tension, suspicion and humour – all positively aroused**. There were also constructs among both a male Part A and Part Bs that alluded to passivity such as grey OK and calm. The sides were indicated by two females, one Part A in reference to 'slow', the other a Part B in reference to 'happy'. **The pelvis was identified by two males, both in relation to aroused emotional constructs – amused and tensed**.

**From the waist down the emotional constructs are mostly negative.** A female Part A attached the word 'slow' to the thighs, and male Part As stated 'scary' and 'restless'. The only Part B (male) to identify the thighs stated 'tensed'. All of these constructs are then connected to the knees, in addition to female Part Bs stating 'tension suspicion' and amusement (positive), and one Part A male stating 'shy'. **The shins were also connected to negative arousal by those who identified them**, in addition to the 'slow'. And the feet were connected to both similar negative notions, in addition to boredom, impatience, alertness and indifference.

*TABLE 7: New elevator emotional constructs with bodily sensations*

| BODY PARTS | Part A female constructs | Part A male constructs | Part B female constructs | Part B male constructs |
|---|---|---|---|---|
| **Head** | Relaxation Pleasant Normal Flowing | Normal Normal Restless Prank-like Anxiety | Comfortable Waiting Interested Cool Awkwardness Relaxed Good Emancipated Easy Calm | Uncomfort Calm Techy Bored Neutral more relaxed |
| **Throat** | Relaxation Normal | Neutral Normal Prank-like | Not relaxed Cool Easy | Uncomfort |
| **Chest** | Neutral Surprised Boredom | Amused Neutral Normal Restless Prank-like Calmness | Mild tension Easy | Mild delight Reserved Uncomfort Uneasy |
| **Shoulders** | | Normal Calmness | Easy | Mild delight Uncomfort |
| **Arms** | Flowing | Normal | Not relaxed | Reserved Neutral more relaxed Amused |
| **Hands** | | Normal | Not relaxed | Reserved |

|  |  | Calmness | Interested<br>Awkwardness | Calm<br>Neutral more relaxed<br>Amused |
|---|---|---|---|---|
| **Stomach** | Flowing<br>Safe | Amused | Interested<br>Good | Amused<br>Mild delight<br>Uneasy |
| **Sides** |  |  |  | Amused<br>Tensed |
| **Pelvis** | Neutral<br>Safe |  |  | Tensed |
| **Thighs** |  | Restless | Mild tension<br>confident | Tensed<br>Uncomfortable |
| **Knees** |  | Restless | Waiting | Tensed<br>Uncomfortable<br>Amused |
| **Shins** | Normal | Normal<br>Restless<br>Prank-like |  | Tensed<br>Uncomfortable<br>Amused |
| **Feet** | Pleasant<br>Normal<br>Flowing | Prank-like | Comfortable<br>Mild tension<br>confident | Tensed<br>Mild delight<br>Amused |

Observing the bodily sensations and emotional constructs results for the new elevator, it is interesting to see that the emotional constructs were mostly connected to the head in the case of female Part Bs, in relation to the new elevator. The constructs were a mixture between aroused and passive, negative and positive emotions. The emotions were dominated by positively passive. Likewise, the male Part As and Bs also featured a mixture of constructs, positive and negative, aroused and passive in relation to sensations in the head. However, the **female Part As only specified positively passive emotions – 'relaxation', 'pleasant', 'normal' and 'flowing'. Interestingly also, the Part As (both female and male) were more positive in relation to the elevator when indicating the throat than the Part Bs.** All of the Part A constructs related to positive passivity, besides one 'prank-like', while Part Bs indicated a range of emotional constructs from 'not relaxed' and 'uncomfort', to 'easy' and 'cool' (both female).

There was a concentration of identification of the **chest area amongst male Part As in connection to the new elevator**. Once again, the emotional constructs attached were mostly positive and humorous emotions. There was one negatively aroused construct, 'restless' stated, the rest were both aroused and passive positive constructs. Among the other participants there was a mixture of emotions articulated in connection with the chest, yet surprisingly not as much among females as among males. The **male Part Bs in particular expressed notions of unease**.

It was mostly males who identified as feeling sensations in their shoulders, equally for Part As as for Part Bs. The Part As emphasised notions of calmness, while the Part Bs arousal both positive and negative. The arms were also interesting in that they were only mentioned twice by Part As (female and male), and in both cases they were connected with 'flowing' and 'normal'. Arms were mentioned mostly by male Part Bs, where they were connected to a mixture of emotions, but notably and in relation to the female Part B who indicated arms, they existed in connection to both negative (not relaxed) and positive (flow) emotions.

The hands were not indicated by female Part As, and were mostly related to negative arousal in female Part Bs, with the exception of one stating 'interested'. Male Part As indicating sensations in the hands gave passively positive constructs, and this was continued among male Part Bs with the exception of one stating 'reserved' and the other stating 'amused'. **Once again, what is interesting to note is the connection between indicating the stomach among males (Part A and B) and the emotion of amusement. All of the females indicating the stomach did this in connection with stating positive constructs**, and only one male Part B indicated the construct 'uneasy'. The sides were only shown by two male Part Bs who connected them to 'amused' and 'tensed' (from the same group). Then the pelvis was indicated twice by female Part As in relation to 'neutral' and 'safe', while the male Part B indicating the pelvis reported 'tensed'.

    The thighs were not identified by female Part As, and were connected to restlessness by male Part Bs. They were interestingly connected to 'mild tension confident' by a female Part B and 'tensed' and 'uncomfortable' by a male Part B. The knees were connected with 'uncomfort' and the negative impacts of time in all cases where they were indicated, apart from one male Part B stating ' amused' and no female Part As identifying them. Two Part As (female and male) connected shins with normality, while the rest connected them with either positive or negative arousal – female Part Bs did not identify the shins.

    The feet seemed to be the site of both negative and positive feelings. All of the female Part As constructs connected to feet were positive, the male Part A was humorous, and two participants (female and male Part Bs) connected feet to tension. So, once again a pattern can be observed in which negativity and arousal were concentrated towards the legs and pelvic region, and excitement related sentiments were shown in the stomach. Positive feelings and sensations of passivity and flow seemed connected to the head.

**Semantic differential results**

The semantic differential results are described here in terms of averages and observable relationships, between the elevators, the participants' roles (i.e. A or B) and gender. The semantic differential method was an efficient way of obtaining information regarding how participants experienced the tested elevators in terms of bodily sensations, elevator design and social emotions. It has also been instrumental in showing the relationship between how people experience these elements first hand, and how others try to evaluate the elevators from another person's point-of-view. It shows where people are more likely to be able to adequately guess another person's evaluations, and where they are less likely to. Please note that the constructs are represented according to their location on the binary scales. That is, words located to the lower-left of the hyphen were at the '1', words at the right were located at the '7'.

*Bodily sensations*

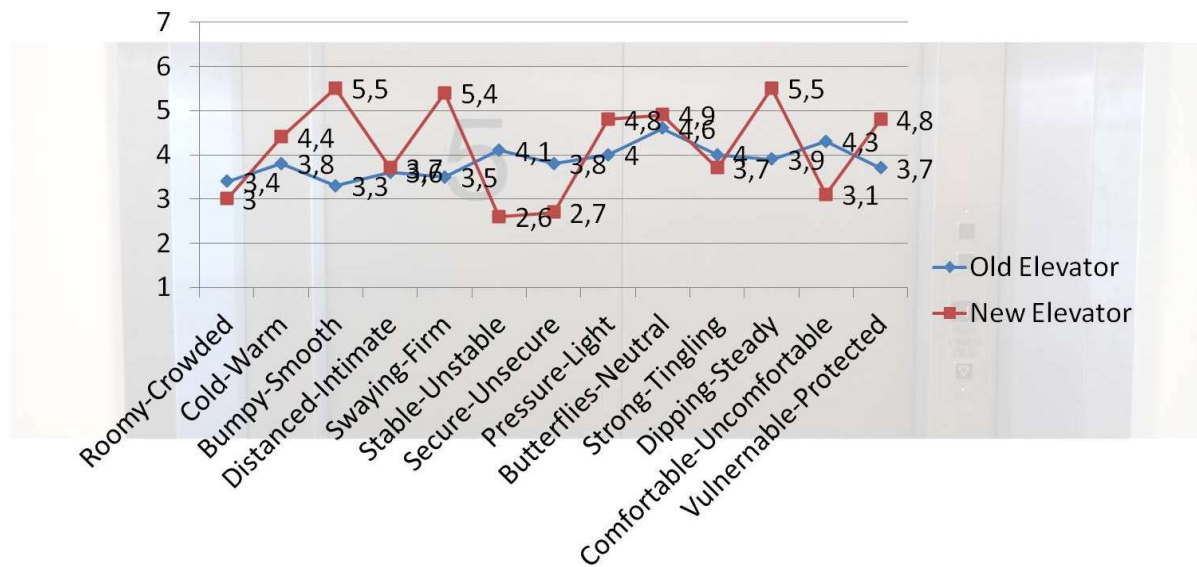

**Fig. 11: Bodily sensation semantic differential total averages**

In light of the semantic differential total averages, it can be seen that the new elevator was experienced as only slightly roomier (3) than the old elevator (3,4). The old elevator (3,8) was experienced as colder than the new elevator (4,4). The **old elevator (3,3) ride was bumpier than the new one (5,5), but then there was an equal experience somewhere in the middle of distancing and intimacy** (3,6 old elevator; 3,7 new elevator), which is connected to how they perceived the sense of space (i.e. roomy-crowded). The new elevator (5,4) was experienced as providing a firm feeling, while the old elevator (3,5) was in the middle (not as firm, yet not swaying). The old elevator (4,1) was experienced as more unstable and the new (2,6) as stable. Also, the new elevator (2,7) was seen as relatively secure while the old elevator was slightly unsecure (3,8). There was no experience of physical pressure in either of the elevators, yet the feeling experienced in the new elevator (4,8) was lighter than in the old elevator (4). In **both elevators participants felt relative neutrality (4,6 old elevator; 4,9 new elevator)** rather than the sensation of butterflies (nervous sensation) in one's stomach. There was also a relatively stable, or neutral sensation experienced in both elevators which slightly went towards a tingling (like goose bumps on the skin = cutis anserina) sensation (3,7 old elevator; 4 new elevator). The new elevator (5,5) was experienced as steadier than the old elevator (3,9), and the new elevator (3,1) was additionally more comfortable than the old (4,3). Participants felt more protected in the new elevator (4,8) than in the old (3,7).

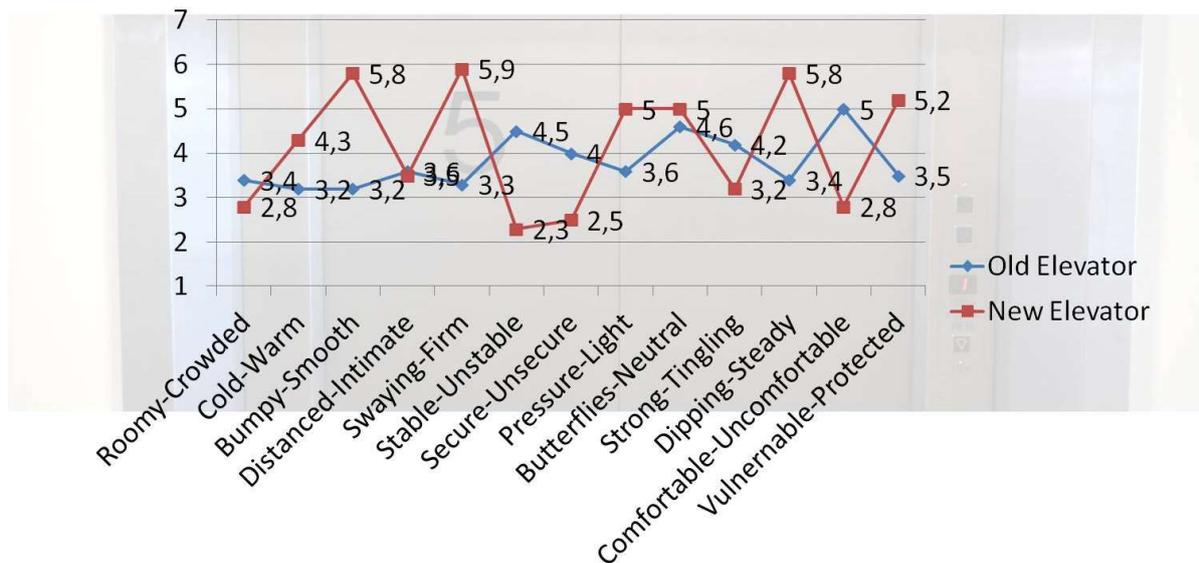

**Fig. 12: Bodily sensation semantic differential Part A averages**

Differences can be seen in the responses when analysing them according to the participant's role. When looking at the chart in general, it can be seen that Part As responses towards the old elevator were much more neutral (between the 3-4 mark) than towards the new elevator. The only stronger responses were regarding **instability** (unstable), **neutrality** (following suit with what is observable), **tingling** sensation (slightly) and **discomfort** (uncomfortable). Responses towards the new elevator were more radical in that based on the results, it can clearly be seen that Part As experienced the new elevator as **roomy**, **somewhat warm**, **smooth**, **firm**, **stable**, **secure**, **light**, **neutral**, **steady**, **comfortable**, and **protected**.

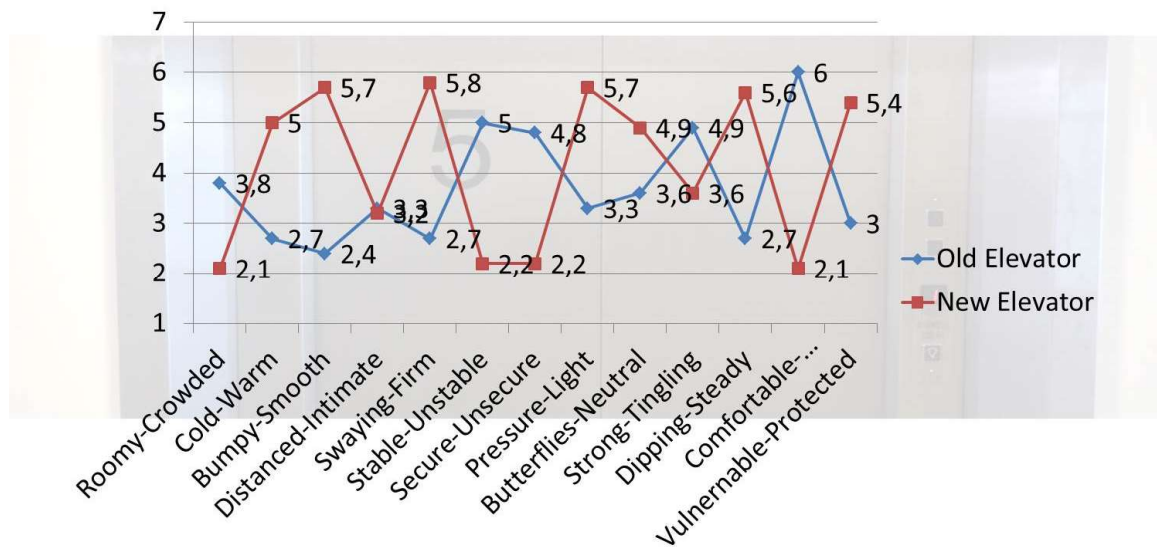

**Fig. 13: Bodily sensations semantic differential Female Part A averages**

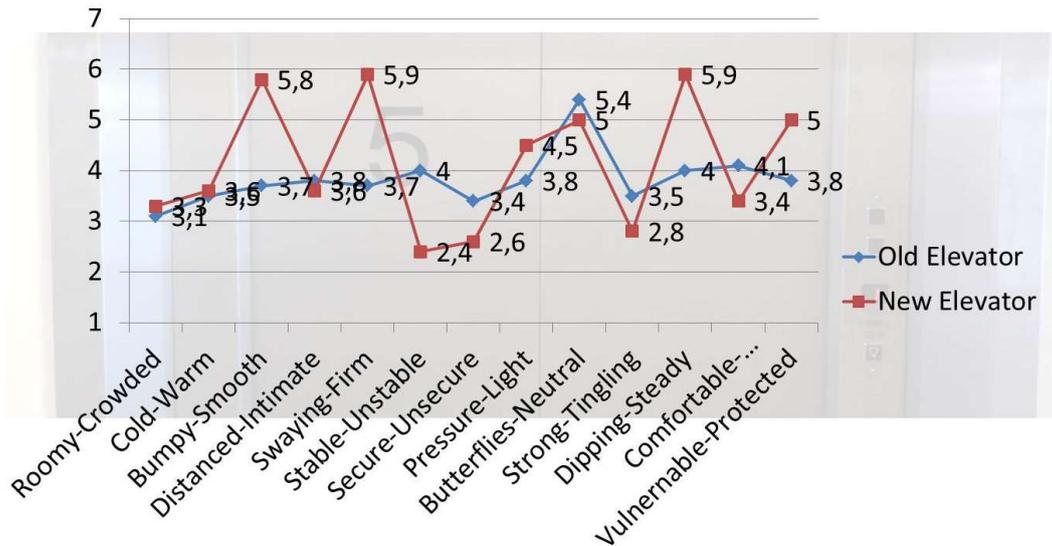

**Fig. 14: Bodily sensations semantic differential Male Part A averages**

When comparing the female Part A responses to the male Part A responses, it can be seen that experiences of the new elevator in comparison to the old varied drastically in the case of the female participants as compared to the male participants. Still, many of the constructs relating to the old elevator were evaluated in terms of neutrality, the only constructs to really stand out being, **cold, bumpy, unstable, unsecure, tingling, dipping, and uncomfortable.** The new elevator was experienced by female Part As as **roomy, warm, smooth, firm, stable, secure, light, neutral, steady, comfortable** and **protected**. The male Part As on the other hand took a yet more neutral line when evaluating the bodily sensations of the old elevator. Here, the only major rise in the chart was at the point of **neutrality**. The male results of the new elevator showed however, that males clearly experienced it to be **smooth, firm, stable, secure, light, neutral, strong, steady** and **protected**. Thus, the new elevator, which was experienced as more positive, received more definite characterisations of the types of bodily experiences induced by the ride. The old elevator was experienced as *neutral*. The differences in male and female participants can be seen particularly in regards to female evaluations of **stability, security and comfort.** Males on the other hand concentrated on **smoothness, stability** and **security**, but without much emphasis on comfort.

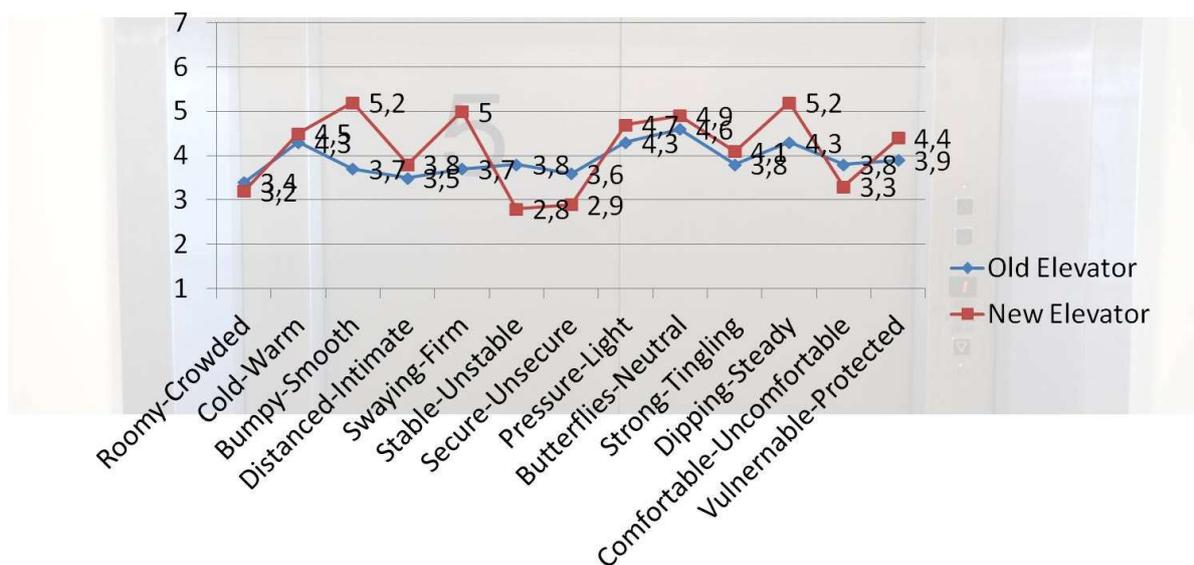

**Fig. 15: Bodily sensations semantic differential Part B averages**

Part Bs seemed to take a more cautious, or neutral line when evaluating the bodily experiences of the elevators on behalf of Part A. Interestingly contrasts can be seen in the case of the 'cold-warm' binary constructs, whereby even the old elevator was experienced as slightly warm. And along these lines the old elevator was perceived as giving a **slightly light bodily sensation, neutrality and steadiness**. The stronger evaluation of constructs once again related to the new elevator, in which Part Bs thought that Part As experienced the elevator as **smooth, firm, stable, secure, light, neutral, strong, steady** and **protected.** Quite similarly to the Part As.

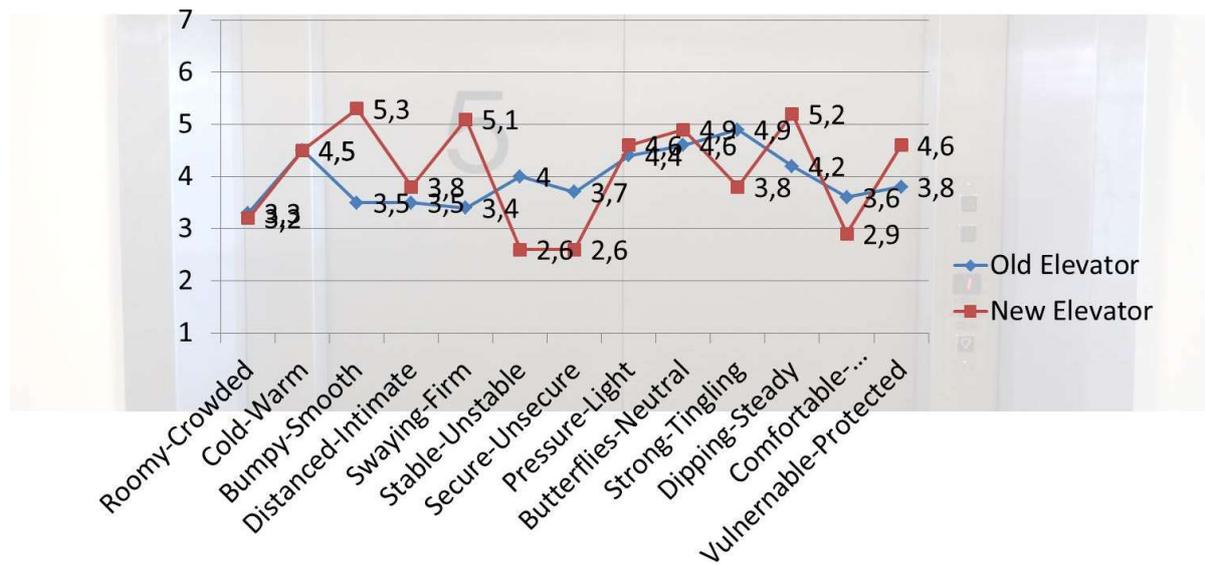

**Fig. 16: Bodily sensations semantic differential Female Part B averages**

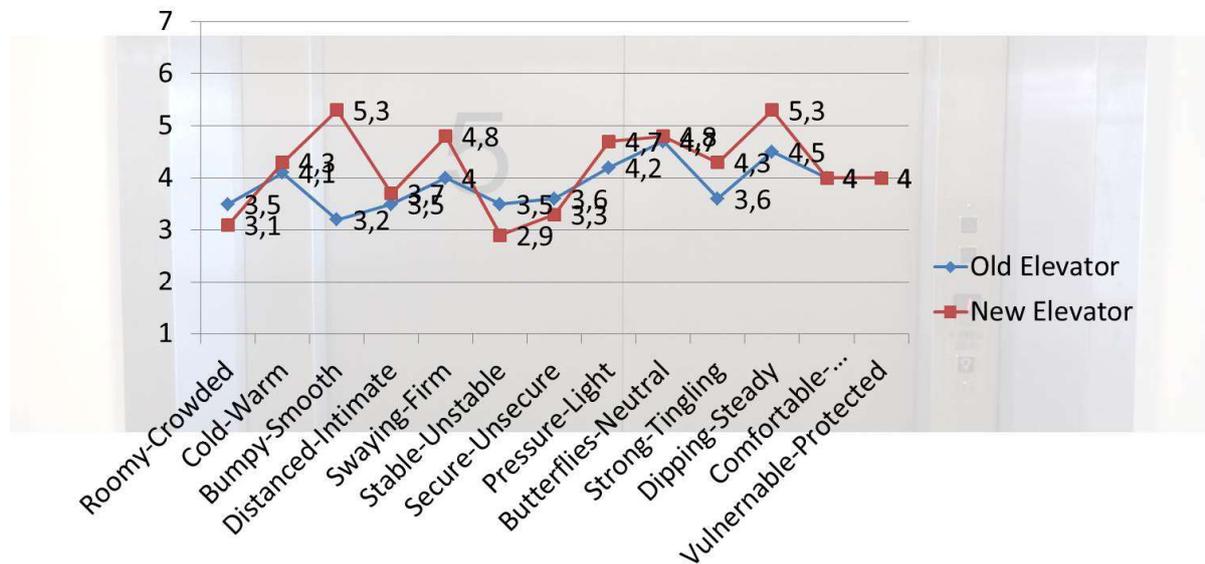

**Fig. 17: Bodily sensations semantic differential Male Part B averages**

The way in which the female and male Part Bs guessed the Part As were experiencing, was surprisingly quite similar. Factors which stand out as being experienced relating to the old elevator were the slight **warmth, neutrality** and **steadiness**. For the new elevator, also there were similarities in it being believed to be perceived as **warm, smooth, firm, secure, light, neutral** and **steady**. Yet, similarly to the results of the Part As, the females also noted the new elevator as being **comfortable**.

*Elevator design*

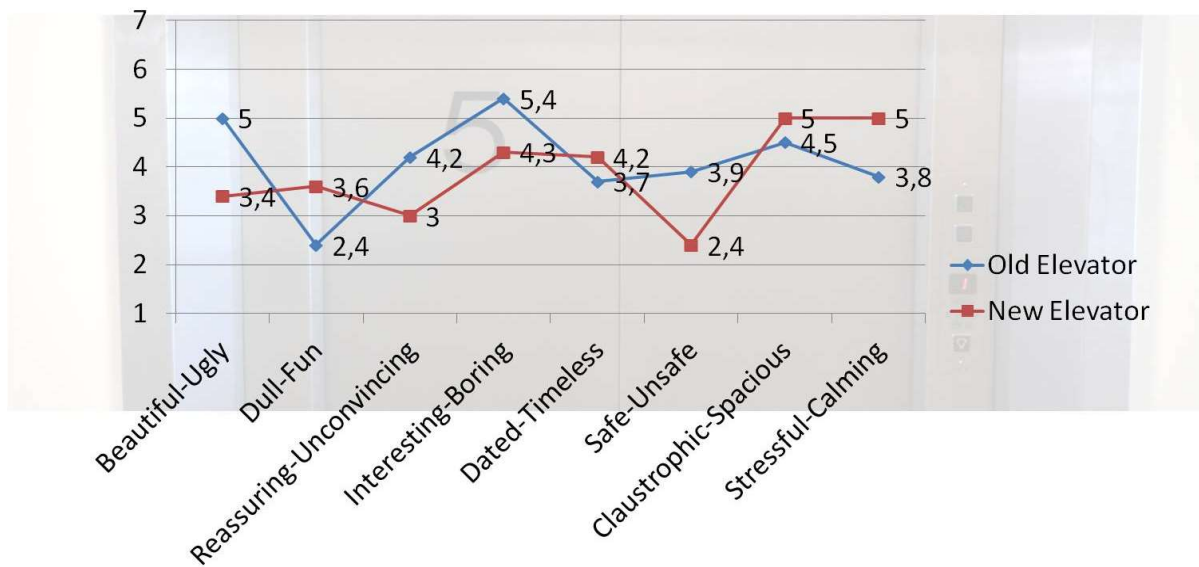

**Fig. 18: Elevator design semantic differential total averages**

The total results of the elevator design semantic differential show that the old elevator was experienced as **ugly** (5), **dull** (2,4), unconvincing (4,2), **boring** (5,4) and spacious (4,5). The new elevator was also considered boring (4,3), timeless (4,2), **safe** (2,4), **spacious** (5) and **calming** (5). Thus, the greatest differences can be seen in terms of ugliness as compared to neutrality and safety.

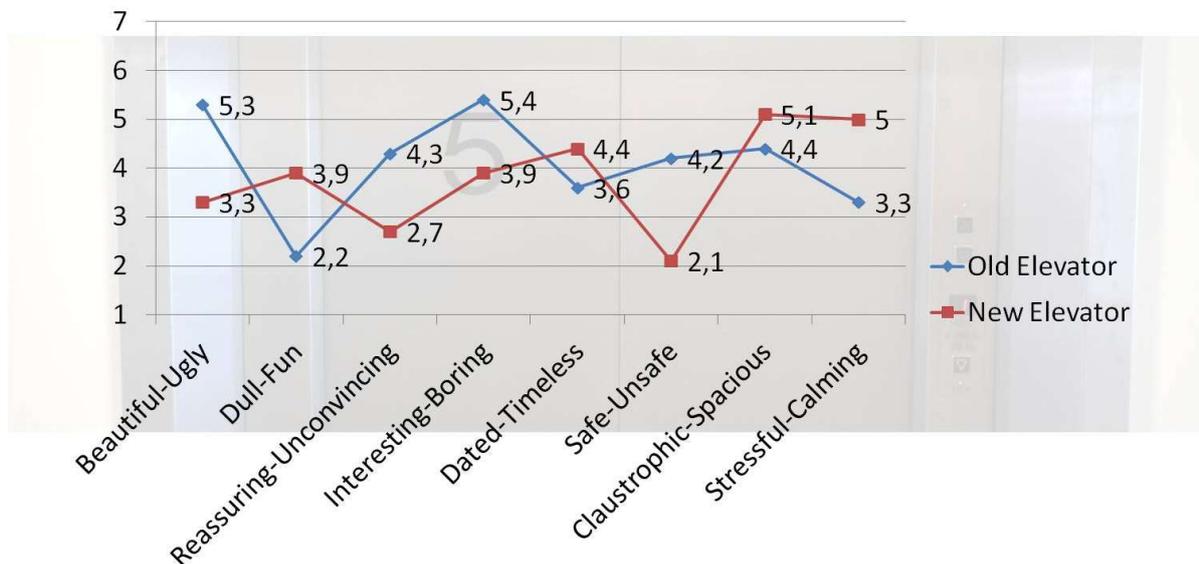

**Fig. 19: Elevator design semantic differential Part A averages**

When looking at the results of Part As, more variations can be seen in terms of the factors of ugliness (old elevator) and neutrality, dullness, unconvincingness and boringness of the old elevator. This is in comparison to the expressed experiences of reassuringness (of the reliability of the elevator), safety, spaciousness and calming elements of the new elevator.

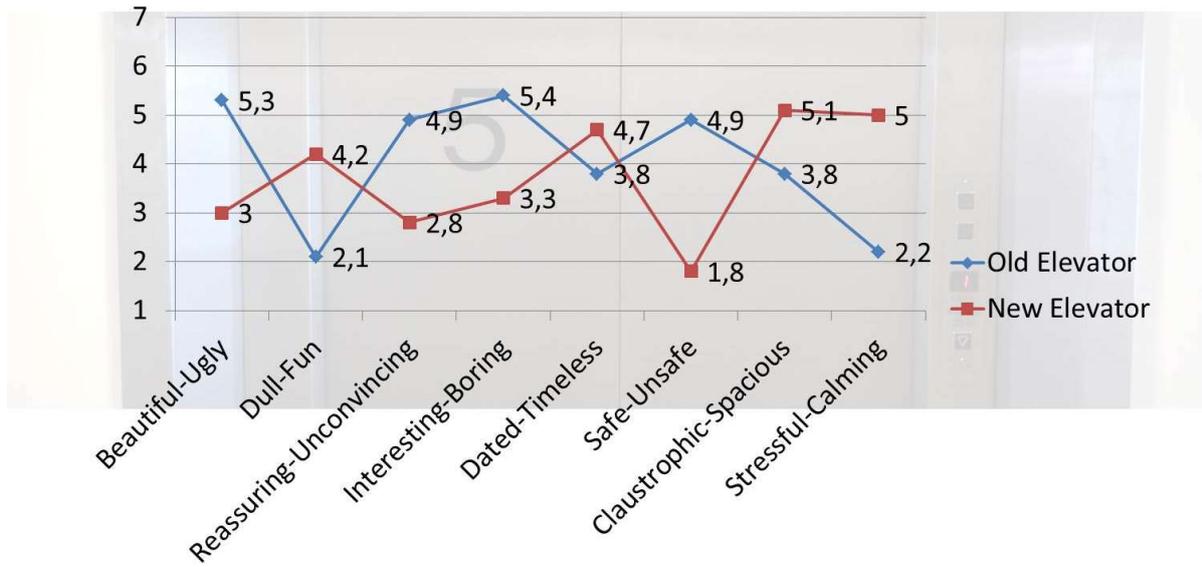

**Fig. 20: Elevator design semantic differential female Part A averages**

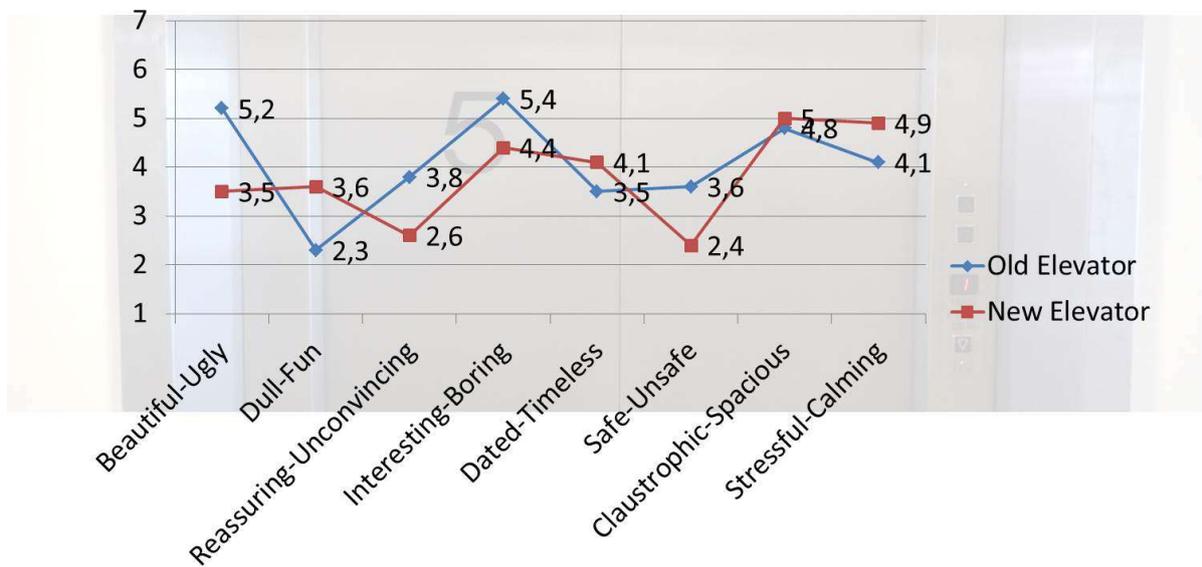

**Fig. 21: Elevator design semantic differential male Part A averages**

There were once again some tendencies among the female attempts to emphasise stronger contrasts between the elevators. The element of **fun** was brought more to the fore by female Part As than male Part As regarding the new elevator. In fact, males expressed the new elevator as well as being slightly **boring**. Also, the **unconvincingness** (unreliability) of the old elevator was more emphasised by females than males so was the feeling of the **lack of safety**, and **stress**. In fact, females and males varied on this note, in that females experienced the old elevator as **stressful**, whereas the males experienced it as slightly **calming**. But for the most part, results were quite similar in that the new elevator was experienced by both genders as being reassuring, timeless, safe, spacious and calming.

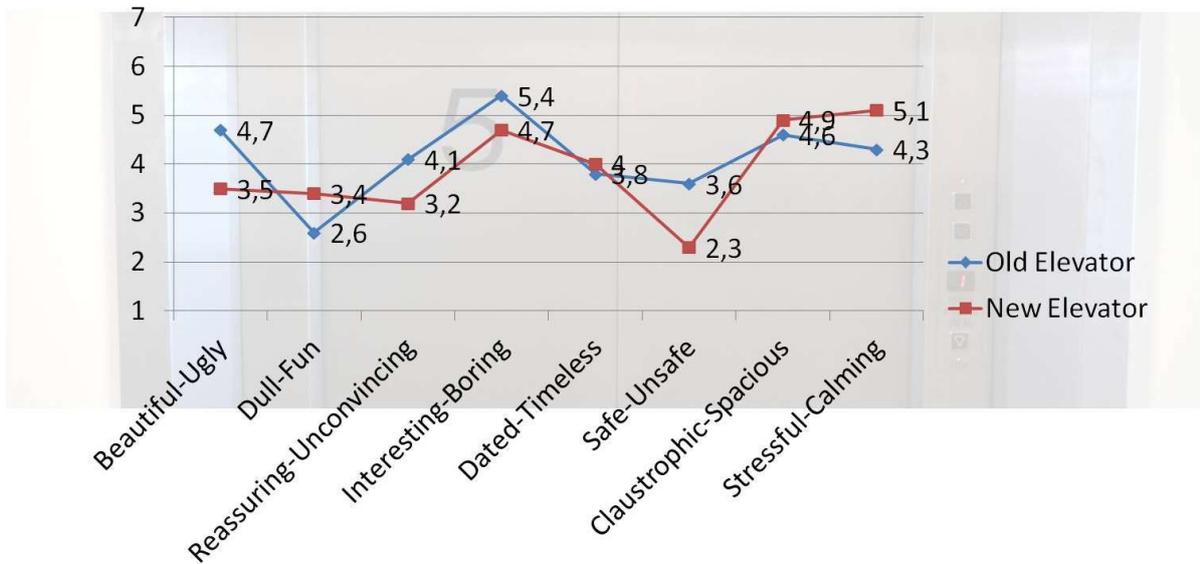

**Fig. 22: Elevator design semantic differential Part B averages**

Part B results proved yet again to exist more along the lines of neutrality, and there was less expressed difference between the elevators. As Part Bs were guessing what Part As were experiencing it can be seen that they felt Part As thought the old elevator was ugly, dull, slightly unconvincing, **boring**, spacious and slightly calming. The new elevator was evaluated as more neutral yet still attracting the evaluations of being **boring, slightly timeless, safe, spacious** and **calming**.

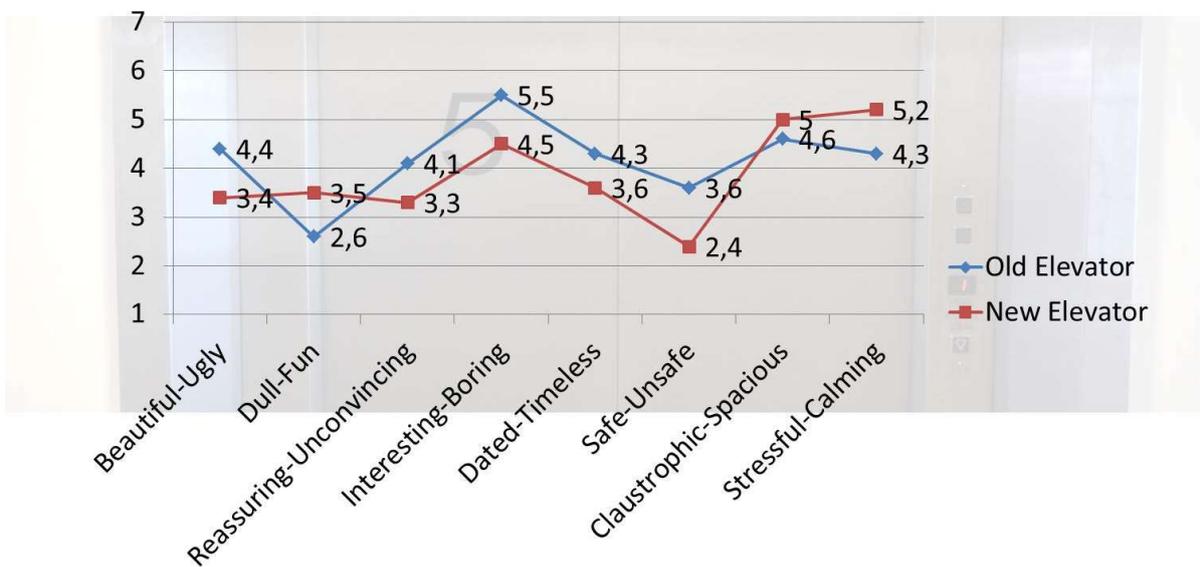

**Fig. 23: Elevator design semantic differential female Part B averages**

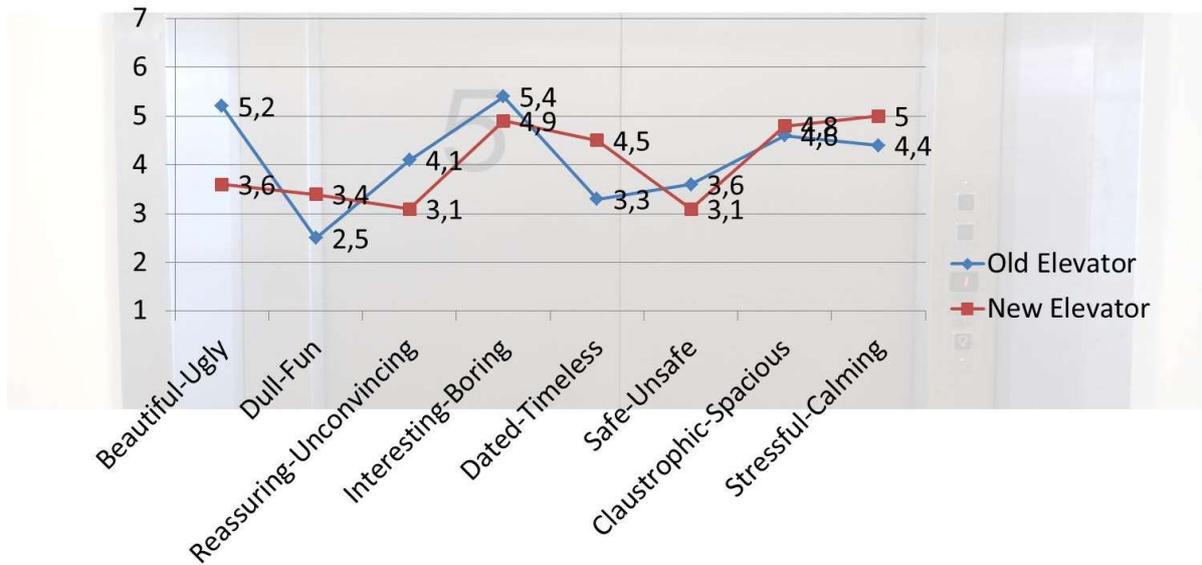

**Fig. 24: Elevator design semantic differential male Part B averages**

The results of female and male Part B evaluations were starkly similar, and somewhat similar to the Part A results. Again, the old elevator was evaluated as **ugly** (stronger response among males), **dull**, **slightly unconvincing, boring, spacious,** and **calming**. Yet, female Part Bs also expressed the old elevator as being **timeless**. Both female and male Part Bs saw the new elevator being experienced as **boring**, yet this time the males evaluated the new elevator as being **timeless** and once again the female Part Bs emphasised the factor of **safety**.

*Social emotions*

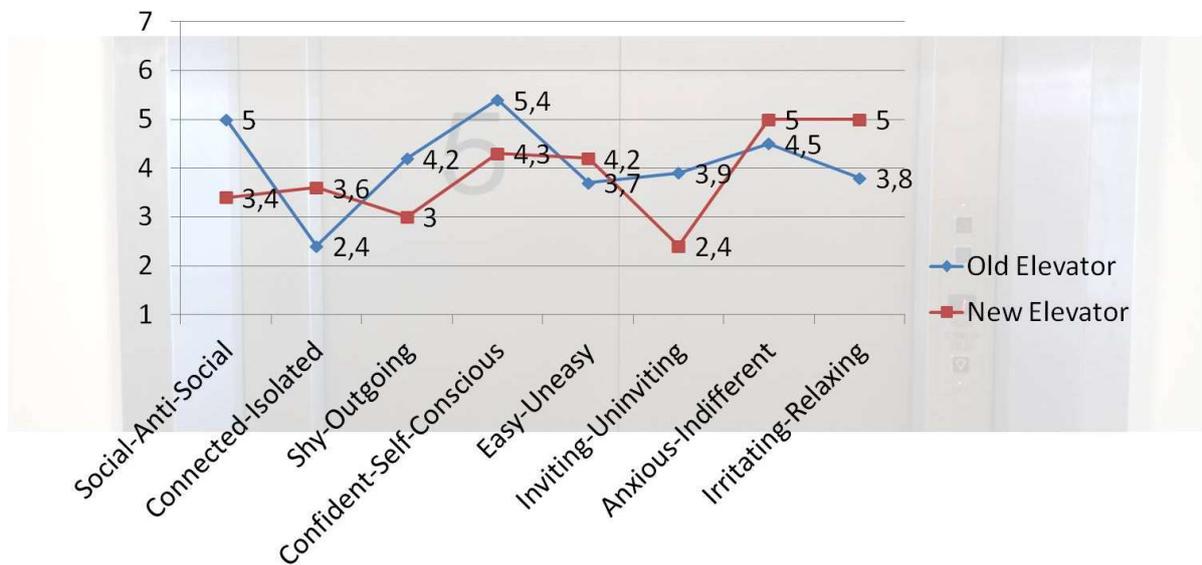

**Fig. 25: Social emotions semantic differential total averages**

This social dimensions component was quite interesting as it was heavily connected to the people with whom the participants undertook the study. Yet, there were still patterns which could be observed in relation to the respective elevators. Due to the social nature of elevator usage, this dimension is important to understand in terms of the impacts induced on the social dynamics on the basis of physical elevator qualities. The old elevator was experienced overall as being **anti-social** (5), yet **connected** (2,4), **outgoing** (4,2)**, self-conscious** (5,4) and **indifferent** (4,5). The social emotions experienced in the new elevator were expressed as slightly **self-conscious** (4,3) and **uneasy** (4,2), **inviting** (2,4), **indifferent** (5) and **relaxing** (5).

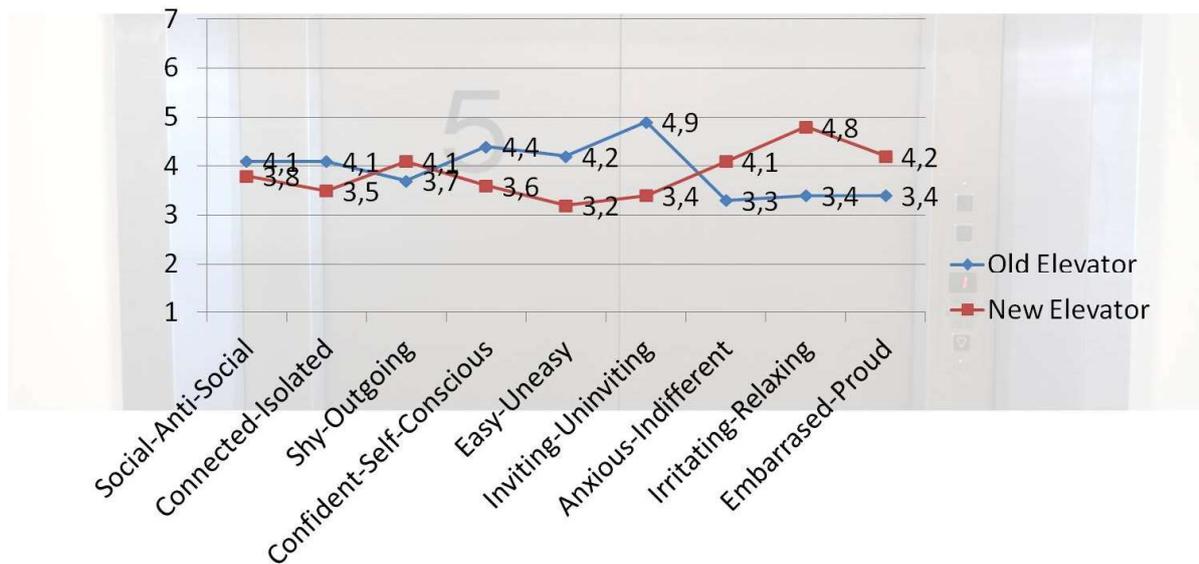

**Fig. 26: Social emotions semantic differential Part A averages**

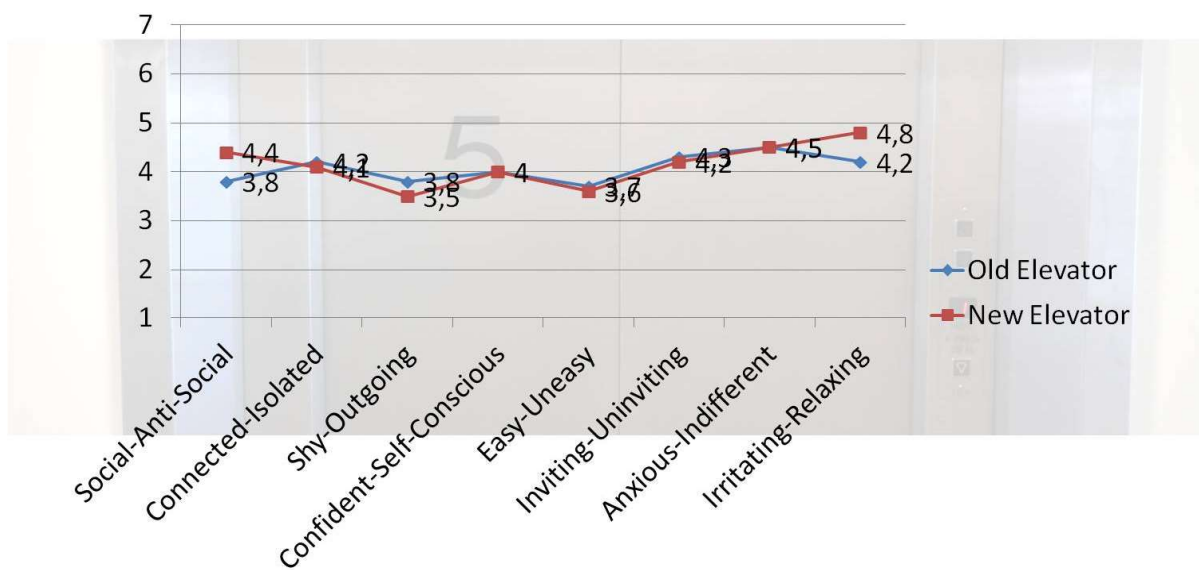

**Fig. 27: Social emotions semantic differential Part A averages**

Here, Part As experienced the old elevator as slightly anti-social, isolated, they felt self-conscious, uneasy, and that it was overall **uninviting**. Whereas, the Part As also felt slightly outgoing, indifferent, **relaxed**, and a sense of pride in the new elevator. Part Bs on the other hand did not vary as greatly in regards to the social emotions they felt Part As had experienced in the respective elevators. Actually, guessed experiences of **anti-social** emotions were attributed to the new elevator, as was the feeling of isolation (slightly), uninvitingness, **indifference** and **relaxation**. In the old elevator Part Bs guessed that Part As felt a slight sense of isolation, unease, **indifference** and to a slightly lesser extent relaxation.

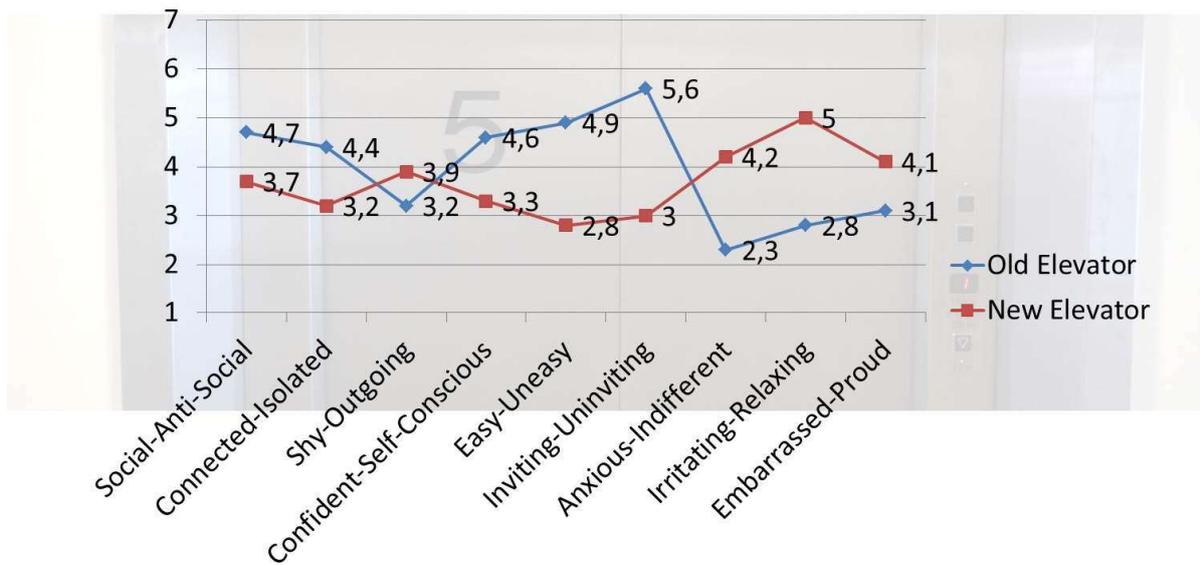

**Fig. 28: Social emotions semantic differential female Part A averages**

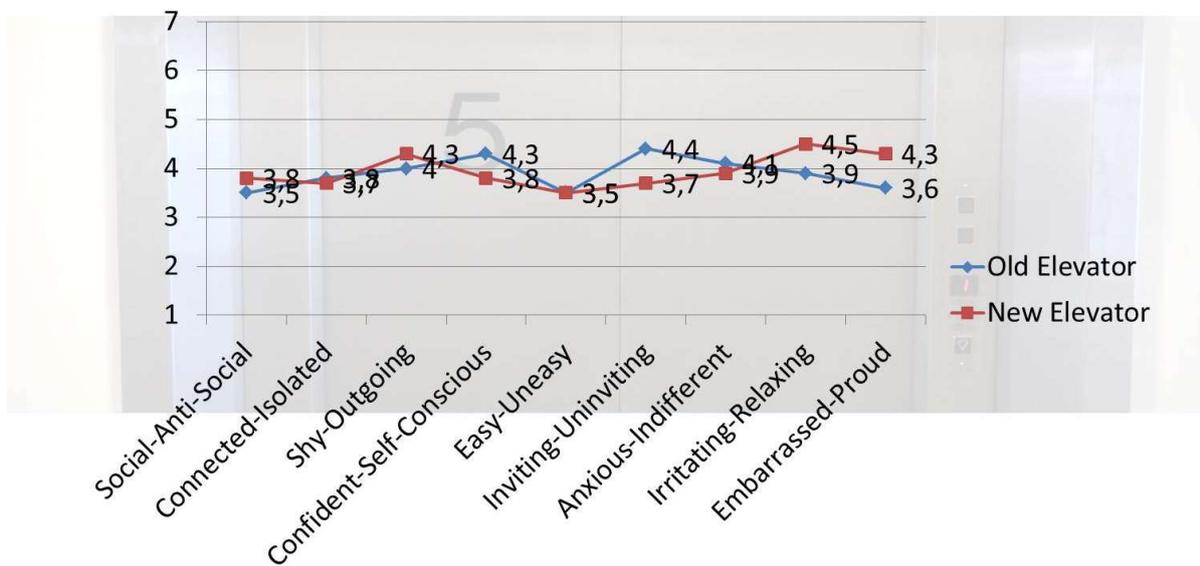

**Fig. 29: Social emotions semantic differential male Part A averages**

Looking at the variations between the female and the male Part As, it can be seen that females rated the old elevator as **anti-social and isolated**, while the males were quite neutral. Males felt slight outgoing in the new elevator, but yet also self-conscious (aware of themselves and possibly concerned that people were looking at them) in the old elevator. Females clearly felt **self-conscious** in the old elevator, as well as **uneasy**, that it was **uninviting** (as did the males), **anxious**, and **irritated**. When asked the 'bonus' Part A question as to whether or not they felt proud or embarrassed (a subconscious reflection of perceiving that one is being watched) both females and males were slightly proud in the new elevator and quite neutral in the old elevator. As can be seen when comparing the graphs, female Part As felt the social emotions on a somewhat stronger level than the male Part As.

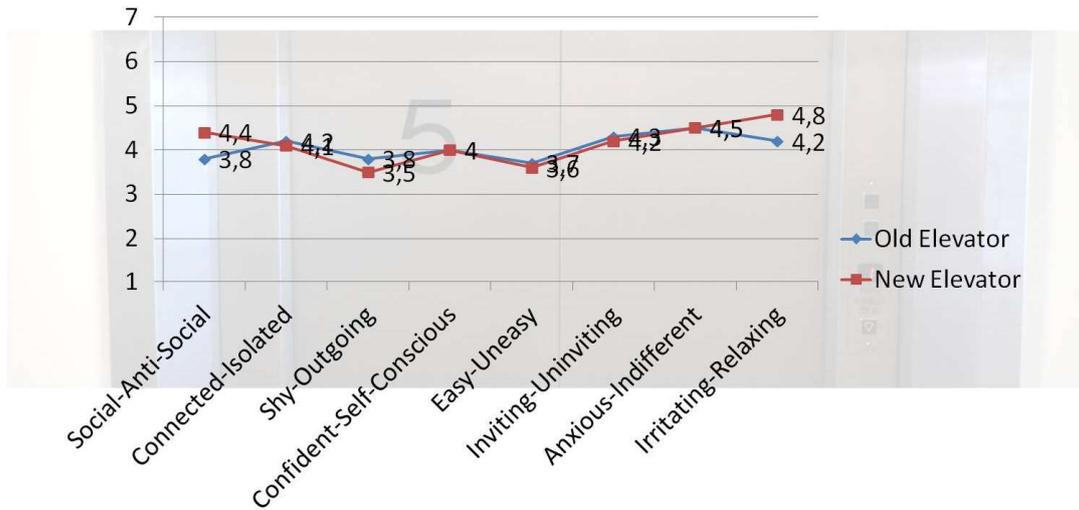

**Fig. 30: Social emotions semantic differential Part B averages**

The overall averages of the social emotions dimension is quite interesting, as on average, Part Bs experienced the new elevator as being slightly ant-social, both elevators as slightly isolating, making them feel self-conscious, they were also shown to be uninviting, in addition to being experienced as indifferent and relaxing.

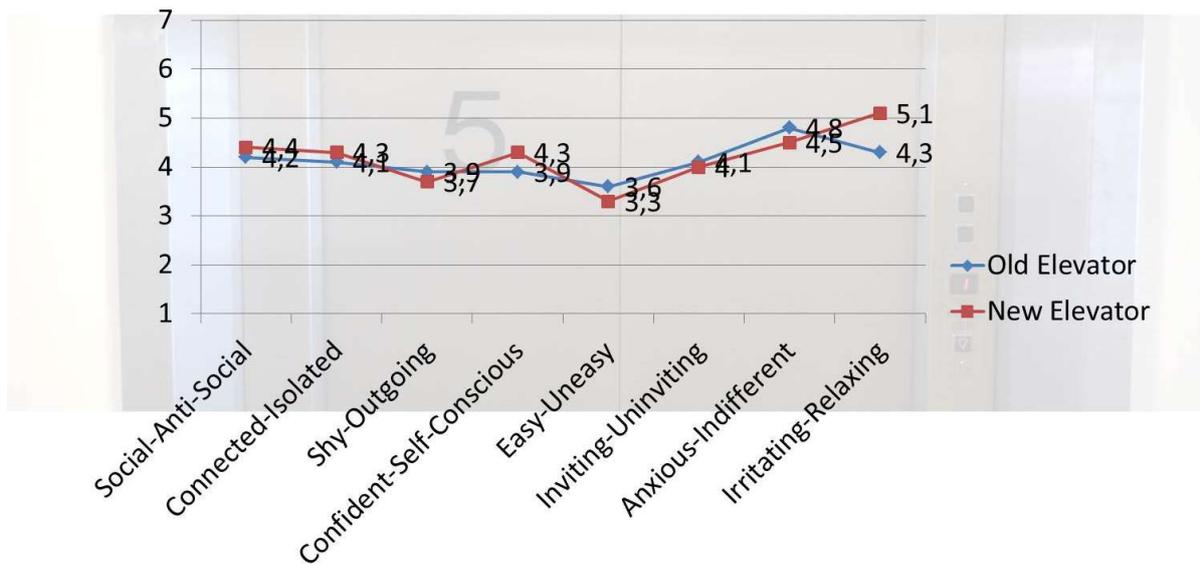

**Fig. 31: Social emotions semantic differential female Part B averages**

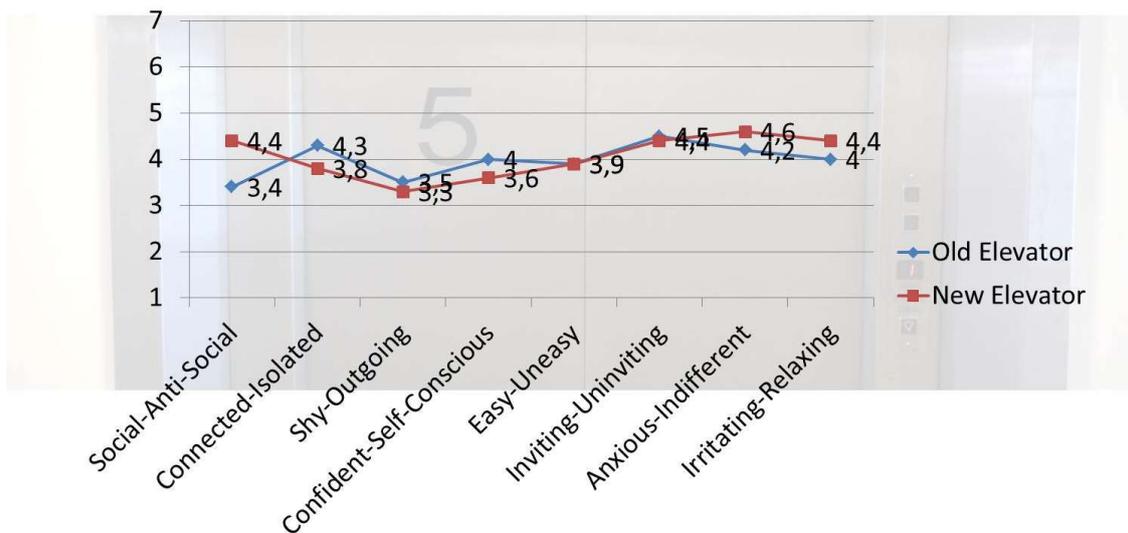

**Fig. 32: Social emotions semantic differential male Part B averages**

Male and female Part Bs gave quite similar responses which were more in line with the male Part Bs than female Part As. Surprisingly, both males and females guessed that the Part As experienced the new elevator as slightly anti-social. There was also the sensing of slight isolation in the new elevator, and they felt that Part As felt slightly self-conscious also in the new elevator. They felt that the new elevator would be perceived as slightly uninviting, yet at the same time that Part As would feel **indifferent** and **relaxed**. The results of the old elevator were similar to that of the new, in that it too was assumed by female part Bs to be experienced as anti-social, yet not by male Part Bs. Both males and females thought the old elevator would be experienced as slightly isolated, uninviting, and again Part Bs predicted that Part As would feel indifferent and relaxed in the old elevator.

While male Part As, as well as both male and female Part Bs assumed a more neutral approach to evaluating the experience of social emotions, female Part As, who were supposed to be evaluating what they themselves felt, had stronger inclinations toward particular constructs. As seen in the results until now, there does not seem to be too much difference, if any between women and men in terms of the ability to read other people's emotions, which means that there might be other instances or *meanings* pertaining to the neurological finding of heightened mirror neuron activity in women as compared to men. But, could this mean also that the mirror neuron activity does not necessarily mean the ability to read others' emotions, rather that women simply *feel* in a stronger way than men? This may explain the female labelling of emotional.

**Levels of difference between Part A and Part B responses**

The following section describes how the female and male Part Bs differed in their semantic differential responses from the Part As. These show the likelihood of accuracy when guessing what another person is experiencing and evaluating qualities on their behalf. The Part Bs had been told to try to *feel* what the Part As were feeling and write their responses according to this. By looking at the total of average point difference between female and male Part B responses, it can be seen that on average, both genders were quite close to predicting or guessing how the Part A was feeling. Despite the visual difference in this graph (please keep in mind that the Y axis starts at 1,5 and ends at 1,86), the actual numerical difference is small. Females were only slightly more accurate in guessing Part A experiences by 0,2 points (the scale was out of 7, meaning that there was the possibility to deviate by 6 points). On average, Part Bs varied from Part As in their responses by 1,7 points.

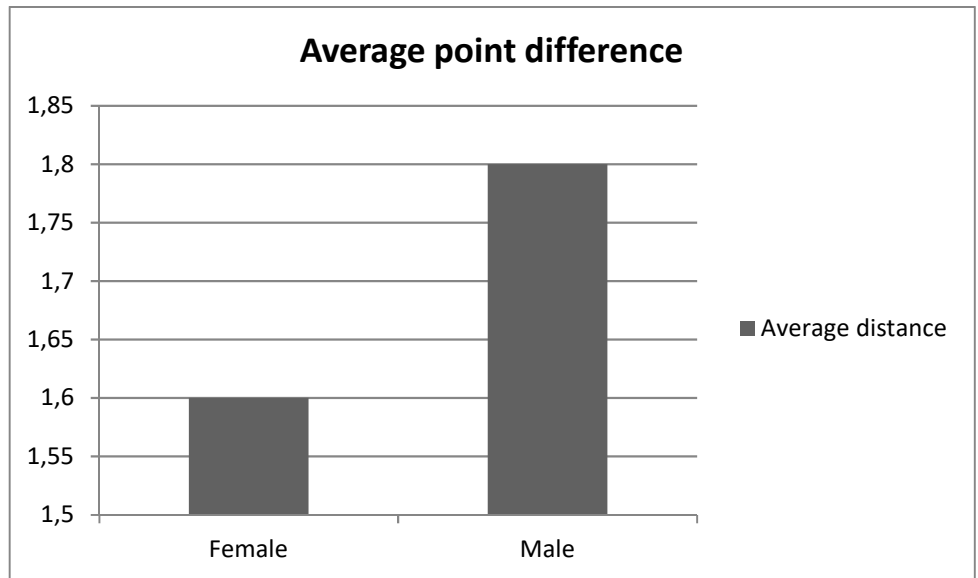

Fig. 33: Average point difference between Part As and Part Bs by gender

When looking at the difference in responses through the dimensions of bodily sensations, elevator design and social emotions, the following can be seen (please note that the order of the binary constructs do not apply in these graphs, they merely describe the factors under evaluation):

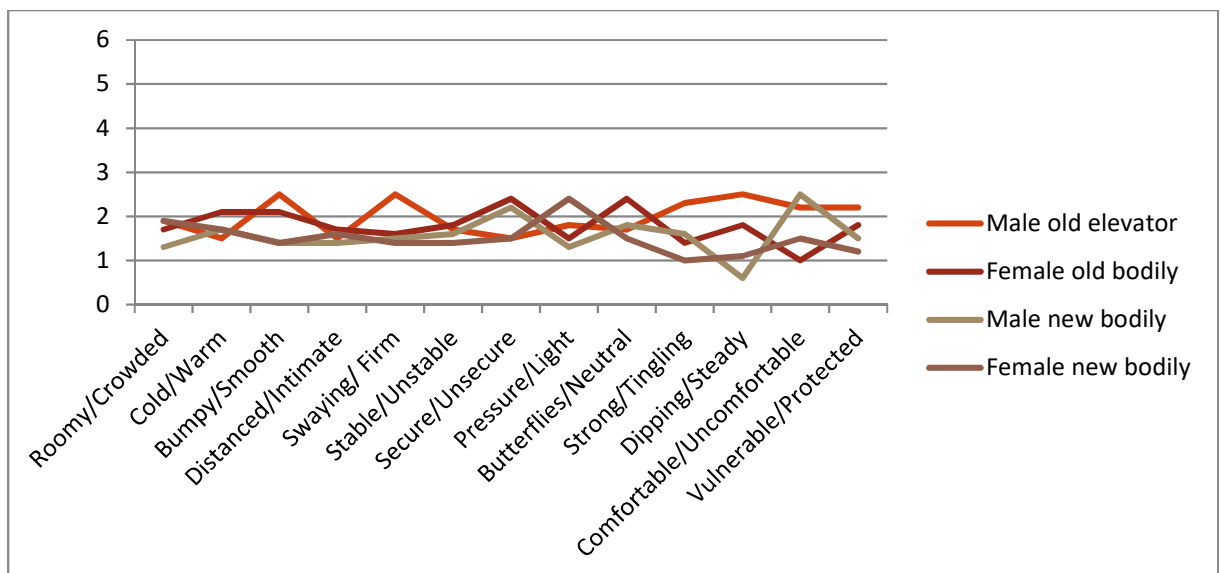

Fig. 34: Levels of difference between Part B and Part A responses regarding bodily experience

While there was not much difference between the genders in their ability to guess what the Part As were experiencing in terms of bodily sensations, females seemed more consistent and able to predict Part As' experiences particularly in relation to the new elevator, and more than the males in the old elevator. Interestingly, once again the factor of **comfort** was more accurately predicted among females than males in both elevators. This held for the sensation of strength/tingling and the pressure level factors. Male Part Bs were particularly accurate when guessing the **steadiness** factor in the new elevator, yet somewhat distanced in the new elevator. The males in particular seemed to vary from the Part As in their responses to the old elevator, especially on the factors of **smoothness, firmness** and this **steadiness** factor. Females had greater accuracy in the old elevator yet varied somewhat in the sensation of **security**, and **neutrality.** The sensation of **pressure** in the new elevator was also somewhat distanced in the female Part B responses. The total average distance in responses towards bodily sensations was **1,7**, females receiving an average of 1,7 and males receiving an average of also 1,7.

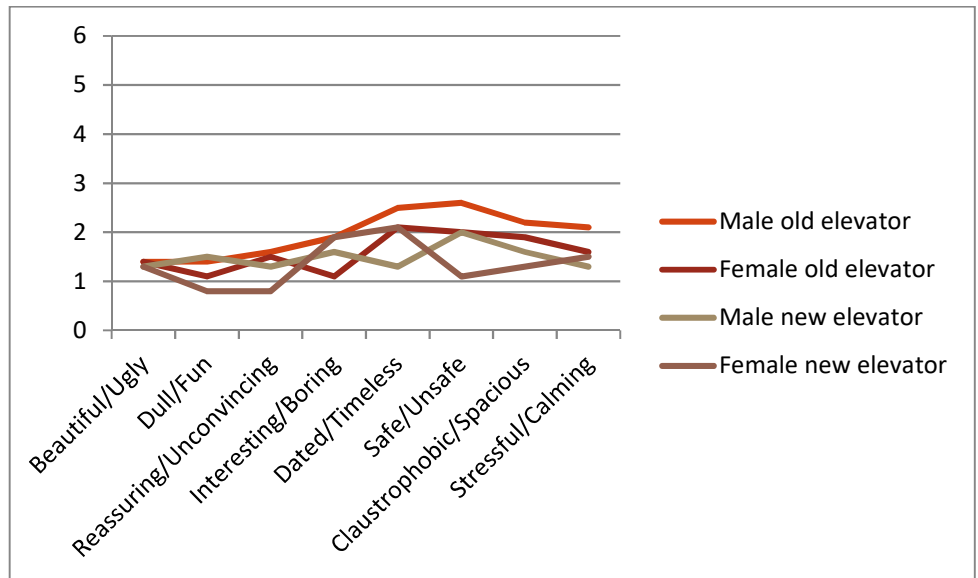

**Fig. 35: Levels of difference between Part B and Part A responses regarding elevator design**

    There was slightly less difference in the responses towards elevator design between Part As and Part Bs, with a total average of 1,6, but also the difference between male and female responses was greater in regards to accuracy. Here, females received an average of 1,5 points difference to the Part As, and males receive 1,8 points difference. So, women on the whole were slightly more accurate in their ability to guess what Part A was thinking about in terms of the physical elevator qualities. In particular, females were quite accurate in their evaluation of the factors of **dull/fun**, **reassuring/unconvincing**, and **safety** of the new elevator. The greatest distance regarding female evaluations related to both the old and new elevator in terms of dated/timelessness. Here, males were more accurate regarding the new elevator, but less accurate regarding the old. For the most part, here it can be seen that males were less able to guess what the Part As were feeling particularly in relation to the old elevator, regarding perceptions of **timelessness** and **safety**. The slightly weaker results of the men regarding experience of safety also continued in the new elevator.

    On the whole, what is interesting to see is that experiences relating to the new elevator were easier to guess than for the old elevator. This may indicate that with newness, smoothness of ride, reliability and attention to interior detail (i.e. polished granite flooring, atmospheric led lighting and carefully finished stainless steel-mirror side panels), the way that people experience is *more likely* to be controlled than if the technologies (elevators) are dated and start working in unpredictable ways.

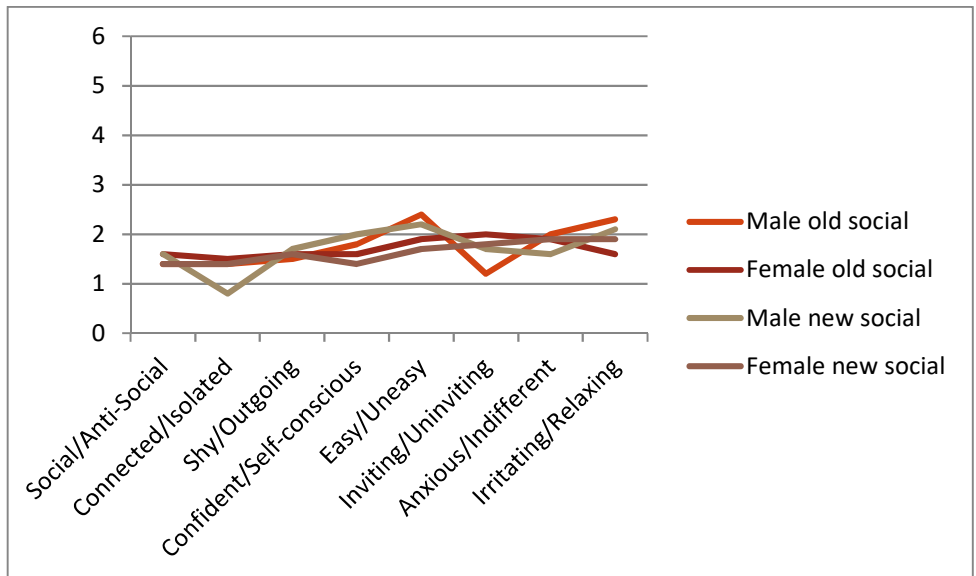

**Fig. 36: Levels of difference between Part B and Part A responses regarding social emotions**

In relation to the dimension of social emotions, a total average difference of 1,7 was achieved. Females and males performed similarly in their ability to guess what Part A what feeling about the social situation. Yet, once again females (1,5 point distance) slight outperformed males (1,8 point distance). The male strong point was their ability to guess the sense of **connectedness** experienced in the new elevator, and the level of **invitingness** in the old elevator. Both in the new and old elevators males were less able to sense the **unease** of Part A, in addition to the level of anxiety/indifference, and **irritation/relaxation**. Female Part Bs quite steadily guessed Part A's experiences within 2 points in both the new and old elevators. They were more aptly able to guess the extent to which Part A experienced social/anti-socialness, as well as connectedness in both the old and new elevators. Also they were quite efficient at guessing the level of irritation that Part A experienced.

All in all, while looking at the results and observing that there was no noticeable difference between the performance of both male and female Part Bs. Rather, a slight tendency for females to be more accurate. However, in all cases it can be seen that guessing how another experiences concrete, physical properties – i.e., of the elevators – is easier than trying to guess what another is feeling, both within their body and in relation to other people.

# Discussion

During the course of the experiments and as seen in the results, some themes and findings did emerge regarding the topic. On the basic level, and as seen through the results – emotional constructs, bodily sensations, and semantic differential – it seems the more stable and consistent the technology, i.e., that of a working and updated elevator, the more uniform the experience (e.g. calmness, relaxation, neutrality). Additionally, by combining the indications of bodily sensations with the emotional constructs it could also be seen that there was a pattern of mostly positive constructs being concentrated towards the head region. Excitement and tension were experienced in the stomach[2]. This corresponds with Nummenmaa et al.'s (2013) results showing that people reported sensations in the stomach and chest area when reporting the non-basic emotion of anxiety. Furthermore, it could be seen in the results of this study that negative feelings were strongly connected to the legs and shins. This somewhat corresponds with Nummenmaa et al.'s (2013) results whereby the basic emotion of fear was experienced in the shins and so too was anger (basic), however the strongest relationship between emotion and bodily sensation in the shins was in the case of happiness (basic).

When viewing the semantic differential results, it can be seen that female Part As – those who were answering from their own perspective – gave the most radical/varied responses, clearly categorising bodily sensations, elevator designs and social emotions in terms of specific characteristics. They also reported more clear contrasts between the elevators. The other participants were not as varied in their results, many times tending towards neutrality, particularly on behalf of the Part Bs. Perhaps Part Bs (both males and females) were weary of making assumptions on behalf of others.

There were also patterns amongst the responses of females and males. Two males reported boringness and restlessness, and there was a tendency in the semantic differential results for the males to rate both the old and the new elevator as boring. Also males rated the old elevator as dull. Females (both Part As and Bs) referred to the 'waiting' aspect of the experience. In their responses, females (both Part As and Bs) more often rated the elevators in terms of comfort and safety in the semantic differential, whereas male Part Bs in particular mentioned the aspect of uncomfort in their qualitative responses. Different social dynamics could be seen to be induced by the different quality elevators. The old elevator tended to induce more self-consciousness (awareness of oneself in relation to the outer world) and was found to be uninviting, while the new elevator showed that people (Part As) reported a slight sense of pride. In the results, it could be seen that the females (both Part As and Bs slightly less) rated the old elevator in terms of being anti-social and isolating. Female Part As rated the old elevator also as stressful.

It can be noticed that the more stable and consistent the elevator experience (new elevator), the more uniform the answers – across participants, quite often referring to the experience in the new elevator as neutral, normal, relaxing and calm. The more inconsistent the experience (the old elevator has a tendency to vary in speed and make random mechanical noises) the less uniform the responses are.

There was the tendency that Part Bs were more accurate in guessing Part As' experiences when basing these experiences on the tangible elevator design characteristics. The 'in-body' characteristics of bodily sensations and social emotions were guessed with less accuracy. Furthermore, Part Bs in general rated the old elevator design in particular as dull,

---

[2] This was an interesting connection in light of Juhani Mattila's (2014) current book which discusses excitement and nervousness in light of fear, as he locates the sense of tension to the stomach.

and females were yet more accurate than men at guessing Part A's experience of the old elevator.

While females only slightly outperformed males in their ability to guess what another person experiences, there may be a relationship between mirror neuron activity and the extent to which people (males and females) experience emotions. As noted earlier, the female Part As had a tendency to be clearer in their expression of differences between the elevators in terms of how they readily characterised their experiences and the design properties, when rating these on the semantic differential scale. However, when expressing their experiences via qualitative emotional constructs (words), female Part As tended to be focused on aspects of time (i.e. the negative experience of waiting in the case of the old elevator) and the positive aspects of neutrality, relaxation (calmness) and safety (in the case of the new elevator).

Male Part As had expressed anxiety both in relation to the old and the new as well as the negative sensation of time. Overall it is interesting to observe the total emotional constructs made by the Part As as compared to Part Bs, as here the Part As seemed to be more focused towards neutrality in their qualitative descriptions only exceeding Part Bs in terms of their negative expressions of time regarding the old elevator. Part Bs on the other hand gave more, either positive or negative constructs regarding negative arousal and positive passivity, in the instance of the old elevator, and positive arousal, negative arousal, negative temporal and positive passivity in the instance of the new elevator. In other words, qualitatively the observers made more distinctions and generalisations than those basing the opinions on their own experiences.

**Yet, quantitatively, females in particular made more distinctions and characterisations based on their own perspectives than those who made observations. Could this relate to our conceptualisation of what qualitative (subjective-based) and quantitative (objective-based) material are?** Could it mean, that in the case of qualitative data, where there is open acknowledgement regarding the subjective and (somewhat) unverifiable nature of qualities in experience, it is easier (or more acceptable) to make generalisations on others' behalf than in the case of quantitative data, whereby we feel that the numbers will always verify some form of truth, which thus means that they *should* be reliable?

# Conclusion

This was an exploratory study, initiated with an interest in discovering the mechanisms of bodily sensations in relation to emotions during elevator travel. For this reason two elevators which contrasted in age, design aesthetics and mechanical fluency were selected. The study sought to implement alternative data collection techniques such as the bodily emotion colouring indication method developed by Nummenmaa et al. (2013), and to see what claims such as the heightened female mirror neuron activity discovery in the neurosciences really meant in terms of elevator experience and the ability to "read minds".

   The results of this study shows that this "mind reading" ability is almost equally as strong in men as in women, but the accuracy of relying on secondary (observational) information to examine the subjective experience of others is roughly 50%. What could be seen however, was the determined nature (less grey areas) of qualitative responses from females who rated their experiences according to semantic differential, and the more distinct categories of qualitative information given by those who were asked to describe another's emotional experiences.

   Based on this small study, it could be seen that the aesthetics in terms of elevator interior design, age and mechanics can influence the ways in which people experience the social dynamics of the elevator, and how much they are aware of their own body and presence inside the elevator. Further, while both genders experienced time negatively, particularly in the case of the old elevator, factors of stress, comfort and safety arose in the semantic differential data from the females – both those who evaluated their own experiences and those who evaluated from the perspectives of others. Male Part Bs mentioned the aspect of uncomfort in their qualitative emotional constructs.

   It would be worthwhile undertaking this study on a larger scale, in a high rise building (office, commercial, recreational etc.) in which the duration of elevator travel time is longer. Further, it should be implemented in conjunction with an ethnographical style study looking at actual use and experiences of everyday elevator travel. By combining these results we may gain more insight into issues such as why the factors of comfort, stress and safety are expressed by females in the highly structured study as compared to the men. And develop deeper insight into mental processes in conjunction with the observable behavioural.